\documentclass[amsmath,amssymb,aps,pra,twocolumn,floatfix,a4paper,superscriptaddress]{revtex4-1}

\usepackage{graphicx}
\usepackage{color,hyperref}
\newcommand{\fref}[1]{Fig.~\ref{#1}}
\newcommand{\Fref}[1]{Figure~\ref{#1}}

\begin{document}

\title{Producing superfluid circulation states using phase imprinting}
\author{Avinash Kumar}
\affiliation{CNRS, UMR 7538, F-93430, Villetaneuse, France}
\affiliation{Universit\'e Paris 13, Sorbonne Paris Cit\'e, Laboratoire de physique des lasers, F-93430, Villetaneuse, France}
\author{Romain Dubessy}
\affiliation{Universit\'e Paris 13, Sorbonne Paris Cit\'e, Laboratoire de physique des lasers, F-93430, Villetaneuse, France}
\affiliation{CNRS, UMR 7538, F-93430, Villetaneuse, France}
\author{Thomas Badr}
\affiliation{CNRS, UMR 7538, F-93430, Villetaneuse, France}
\affiliation{Universit\'e Paris 13, Sorbonne Paris Cit\'e, Laboratoire de physique des lasers, F-93430, Villetaneuse, France}
\author{Camilla De Rossi}
\affiliation{Universit\'e Paris 13, Sorbonne Paris Cit\'e, Laboratoire de physique des lasers, F-93430, Villetaneuse, France}
\affiliation{CNRS, UMR 7538, F-93430, Villetaneuse, France}
\author{Mathieu de Go{\"e}r de Herve}
\affiliation{Universit\'e Paris 13, Sorbonne Paris Cit\'e, Laboratoire de physique des lasers, F-93430, Villetaneuse, France}
\affiliation{CNRS, UMR 7538, F-93430, Villetaneuse, France}
\author{Laurent Longchambon}
\affiliation{Universit\'e Paris 13, Sorbonne Paris Cit\'e, Laboratoire de physique des lasers, F-93430, Villetaneuse, France}
\affiliation{CNRS, UMR 7538, F-93430, Villetaneuse, France}
\author{H{\'e}l{\`e}ne Perrin}
\affiliation{CNRS, UMR 7538, F-93430, Villetaneuse, France}
\affiliation{Universit\'e Paris 13, Sorbonne Paris Cit\'e, Laboratoire de physique des lasers, F-93430, Villetaneuse, France}

\begin{abstract}
We propose a method to prepare states of given quantized circulation in annular Bose-Einstein condensates (BEC) confined in a ring trap using the method of phase imprinting without relying on a two-photon angular momentum transfer. The desired phase profile is imprinted on the atomic wave function using a short light pulse with a tailored intensity pattern generated with a spatial light modulator. We demonstrate the realization of ``helicoidal'' intensity profiles suitable for this purpose. Due to the diffraction limit, the theoretical steplike intensity profile is not achievable in practice. We investigate the effect of imprinting an intensity profile smoothed by a finite optical resolution onto the annular BEC with a numerical simulation of the time-dependent Gross-Pitaevskii equation. This allows us to optimize the intensity pattern for a given target circulation to compensate for the limited resolution.
\end{abstract}
\maketitle

\section{\label{sec:introduction}Introduction}
In recent years, the variety of confinement potentials available for trapping ultra cold atoms has developed dramatically. Beyond harmonic potentials, the use of dipole traps \cite{Grimm2000}, magnetic traps, adiabatic potentials \cite{Garraway2016,Perrin2017}, or a combination of them has given access to a wide range of geometries, including optical lattices \cite{Bloch2005}, low-dimensional confinement \cite{QGLD2003}, box traps for uniform gases \cite{Gaunt2013}, narrow channels between reservoirs \cite{Brantut2012}, ring traps \cite{Morizot2006,Heathcote2008,Ryu2007}, and other arbitrary trap shapes \cite{Corman2014,Aidelsburger2017}. Taking advantage of these new tailored potentials, quantum transport experiments with quantum gases have been carried out \cite{Brantut2012,Eckel2014a,Eckel2016,Ryu2015,Caliga2016}, with strong analogies between quantum gas setups and mesoscopic condensed matter devices \cite{AmicoAtomtronics}.

In particular, annular quantum gases confined in ring traps can sustain persistent flows with a quantized circulation \cite{Ryu2007,Ramanathan2011,Moulder2012}, which are analogous to persistent currents in superconducting rings with a quantized magnetic flux \cite{Akkermans2007}. In such a state, the condensate wave function presents a phase winding $2\pi\ell$ around the ring, giving rise to a quantized circulation $\ell h/m$, where $m$ is the atomic mass, $h$ the Planck constant, and $\ell\in\mathbb{Z}$ the winding number. Circulation states have been studied in the presence of a focused laser spot providing a rotating potential barrier, yielding a weak link along the ring in the spirit of superconducting quantum interference devices \cite{Ryu2013,Eckel2014a}.

The circulation state can be prepared in different ways. First, a potential barrier localized within the ring, produced for example by a focused laser beam, and rotated fast enough can excite the quantum gas and let vortices penetrate through the barrier, producing in turn a circulating state \cite{Piazza2009,Wright2013a}. While this technique has proven its efficiency in the preparation of circulation states with a well-defined winding number $\ell$ \cite{Wright2013a}, it remains limited to relatively small values of $\ell$ and necessitates a long preparation time, which can be an issue if the lifetime of the sample is limited or if fast operations on the wave function are required for quantum information protocols.

Another successfully demonstrated method relies on the direct imprint of a given phase winding with the winding number $\ell$ onto the condensate wave function. This has been achieved by two-photon Raman transfer, one of the laser beams being a Laguerre-Gauss beam carrying an orbital angular momentum with a helicoidal phase \cite{Andersen2006,Moulder2012}. The duration of the Raman pulse is on the order of a few microseconds, which makes this method very efficient for the fast preparation of a given circulation state, determined by the order $\ell$ of the Laguerre-Gauss mode. However, this technique makes use of the internal structure of the atomic ground state, coupling different Zeeman substates, and makes difficult its application to atoms confined in a magnetic trap.

In this paper we propose a phase imprinting method using a pulsed light shift potential with a tailored helicoidal intensity profile, where the light intensity varies linearly with the azimuthal angle $\theta$, but with no topological charge \cite{Dobrek1999,Swingle2005}. 
Zheng and Javanainen \cite{Zheng2003} have studied the effect of phase imprinting a realistic phase profile, with a finite light intensity gradient after a loop, on a one-dimensional annular gas. These authors found that phase imprinting alone was not able to create a well-defined circulation state if the gas is confined in a rotationally invariant ring trap. Here we show through numerical simulation of the Gross-Pitaevskii equation that breaking the rotational invariance with a localized potential barrier allows one to circumvent this issue and to establish a controlled circulation by phase imprinting. Moreover, this method can be faster than the stirring method and insensitive to the magnetic sublevels, which makes it applicable to atoms confined in magnetic potentials. Since the intensity profile can be engineered to any pattern, this technique is more versatile and its scope can be extended to the preparation of other target states, beyond circulation states.

The presentation of this work is organized as follows. In Sec.~\ref{sec:PhaseImprinting} we present the principle of phase imprinting and show how to implement experimentally an helicoidal intensity profile with a spatial light modulator (SLM). In Sec.~\ref{sec:Simulations} we explore the effect on the atomic dynamic of the phase imprint of a realistic intensity profile with a finite resolution. This allows us to optimize the phase profile for reaching a given circulation state. Section \ref{sec:Dissipation} shows the effect of dissipation in reaching the steady state. Finally we present our conclusions in the last section.

\section{\label{sec:PhaseImprinting}Phase imprinting}
Phase imprinting is an effective technique to induce a given dynamics in a Bose-Einstein condensate \cite{Burger1999,Denschlag2000,Moulder2012}. Two approaches have been demonstrated to design the phase of the wave function with a given space dependence. First, the phase can be imparted by a Raman two photon process, the phase or angular momentum carried by the photon then being imprinted onto the atomic wavefunction, giving rise to an induced atomic momentum \cite{Kasevich1991}, an angular momentum \cite{Moulder2012}, or both \cite{Ryu2007}. Alternatively, when a spatially dependent potential is pulsed for a time short as compared to the time for atomic motion (e.g. the trap period), the potential is merely imprinted on the atomic phase. This potential can be conveniently produced by a far off-resonance, tailored laser pulse, as demonstrated for instance for the preparation of a soliton \cite{Denschlag2000}. In the present work we follow the second approach to imprint an arbitrary phase.

\subsection{Principle of phase imprinting\label{sec:principle}}
We start with a gas in its ground state, described within the mean-field limit by the normalized wave function $\psi_{0}$. A far-off-resonance light beam, with a two-dimensional (2D) intensity profile $I(x,y)$, is then pulsed onto the atoms, which gives rise to the light shift potential $U(x,y)=\alpha I(x,y)$ proportional to the local light intensity, with $\alpha$ being a factor proportional to the polarizability, which is given in the two level approximation by \cite{CohenDGOVO}
\begin{equation}
\alpha=\frac{\Gamma}{\Delta}\frac{\hbar\Gamma}{8I_s}.
\end{equation}
Here $\Delta$ is the detuning of the light field from the atomic resonance, $\Gamma$ is the transition line width, and $I_s$ is the saturation intensity. If such a potential is pulsed for a time duration $\tau$, much smaller than the time scales set by the trapping frequencies of the condensate, the wave function after the pulse is given by
\begin{equation}\label{eq:PhaseImprinting}
\psi(x,y,\tau) = e^{-\frac{i}{\hbar} U(x,y)\tau}  \psi_{0}(x,y).
\end{equation} 
Hence in the limit of small $\tau$, the potential will simply add the phase $\varphi(x,y)=-U(x,y) \tau/ \hbar$ to the ground-state wave function $\psi_{0}$.

\begin{figure}[t]
\centering
\begin{tabular}{ccc}
\multicolumn{3}{c}{\includegraphics[width=\columnwidth,clip=true]{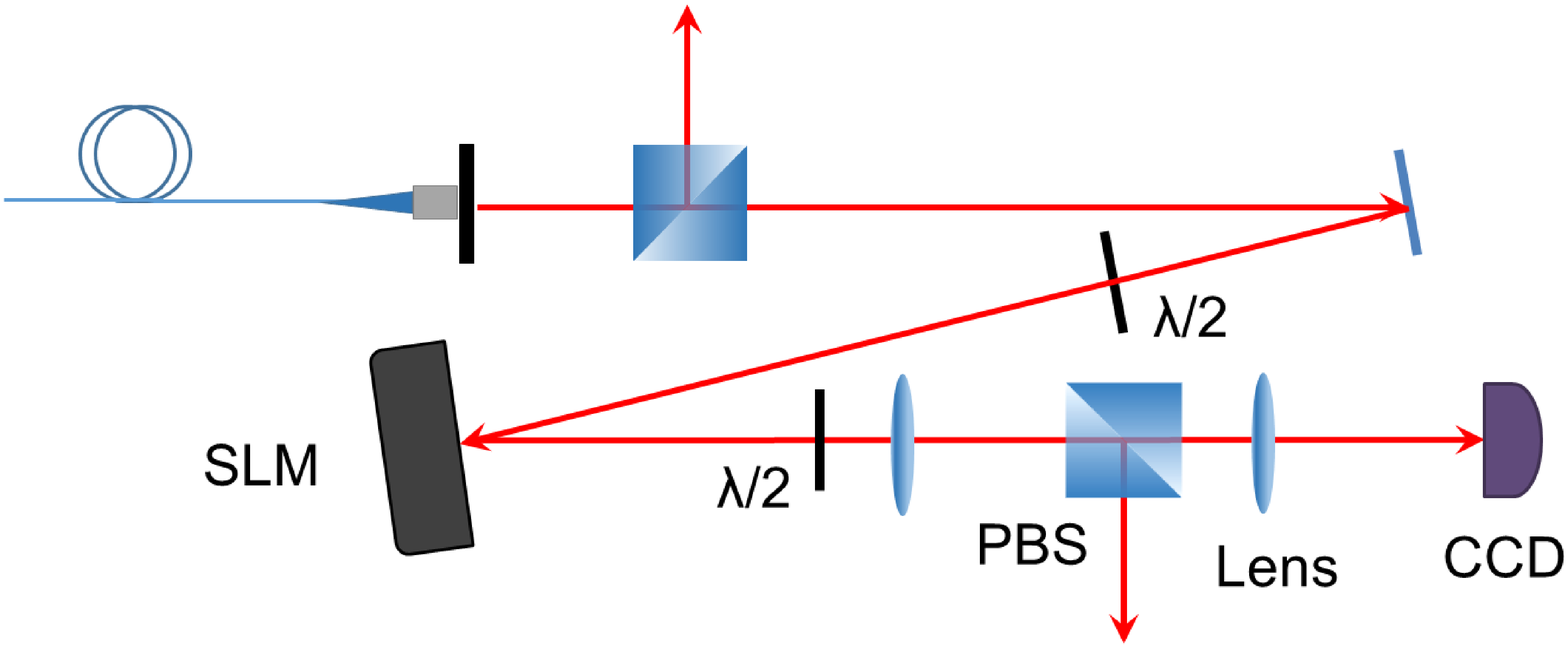}}\\
\multicolumn{3}{c}{(a)}\\[0.05\columnwidth]
\includegraphics[height=0.25\columnwidth,clip=true]{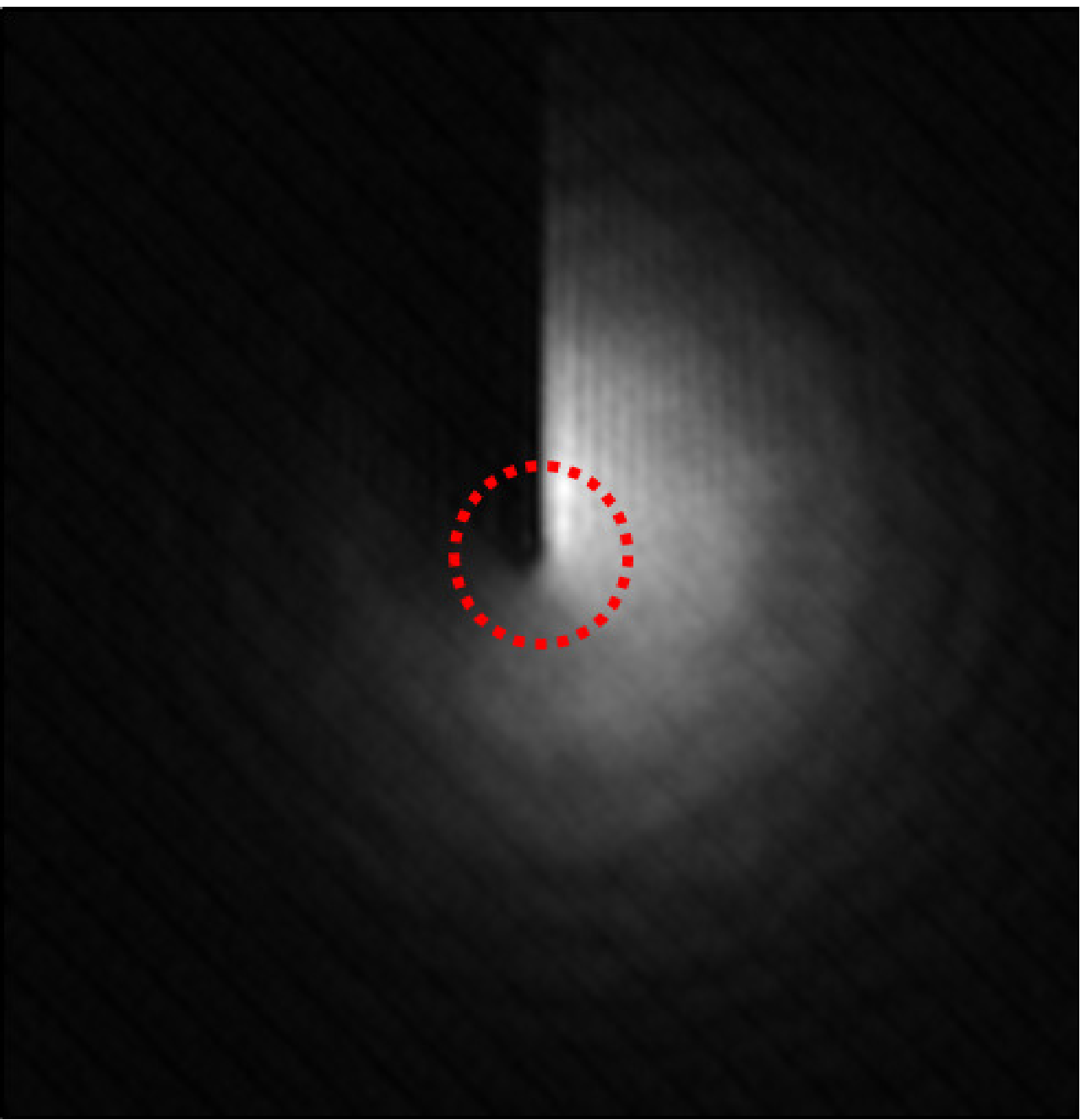} &\includegraphics[height=0.25\columnwidth,clip=true]{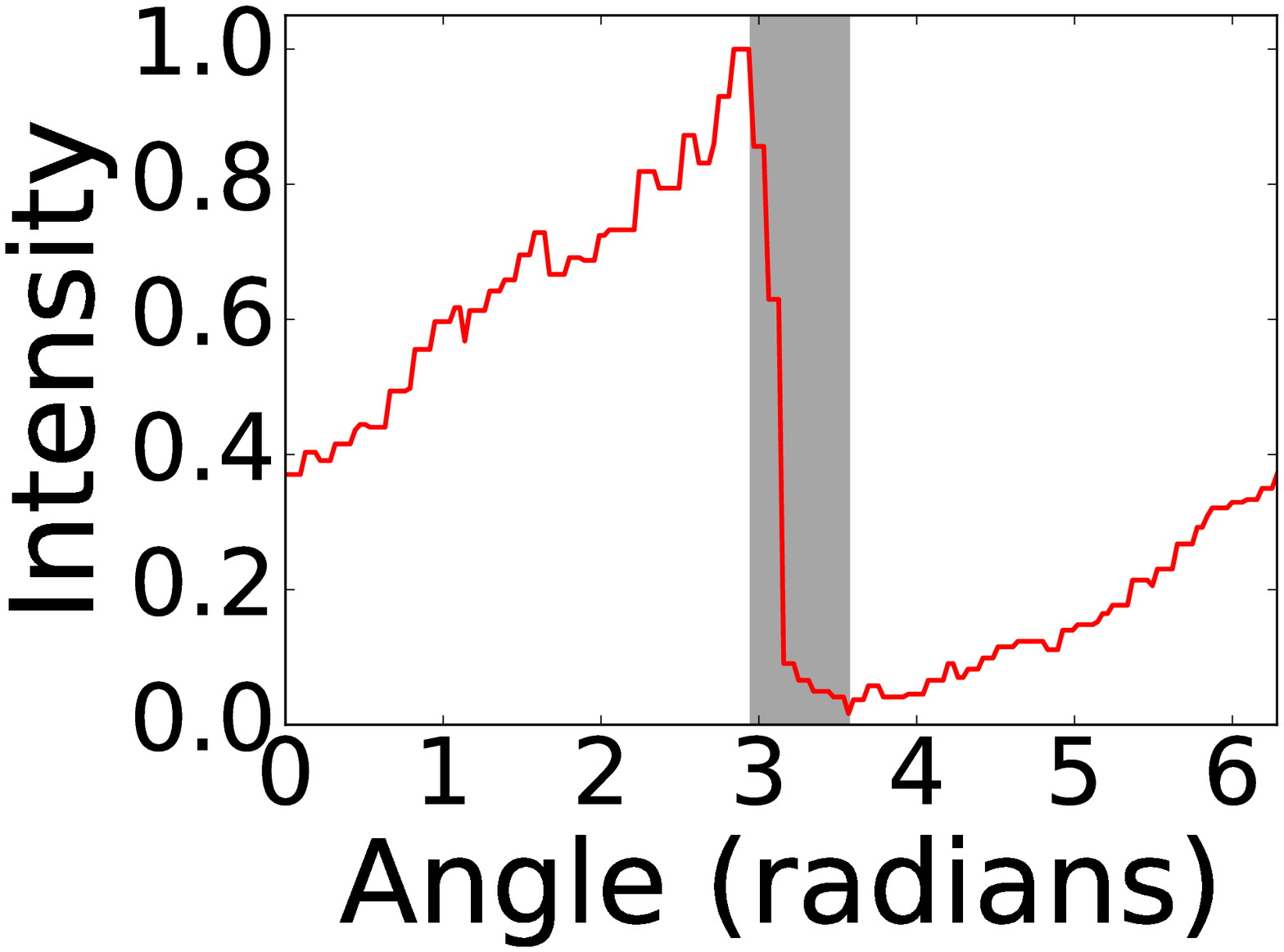} &\includegraphics[height=0.25\columnwidth,clip=true]{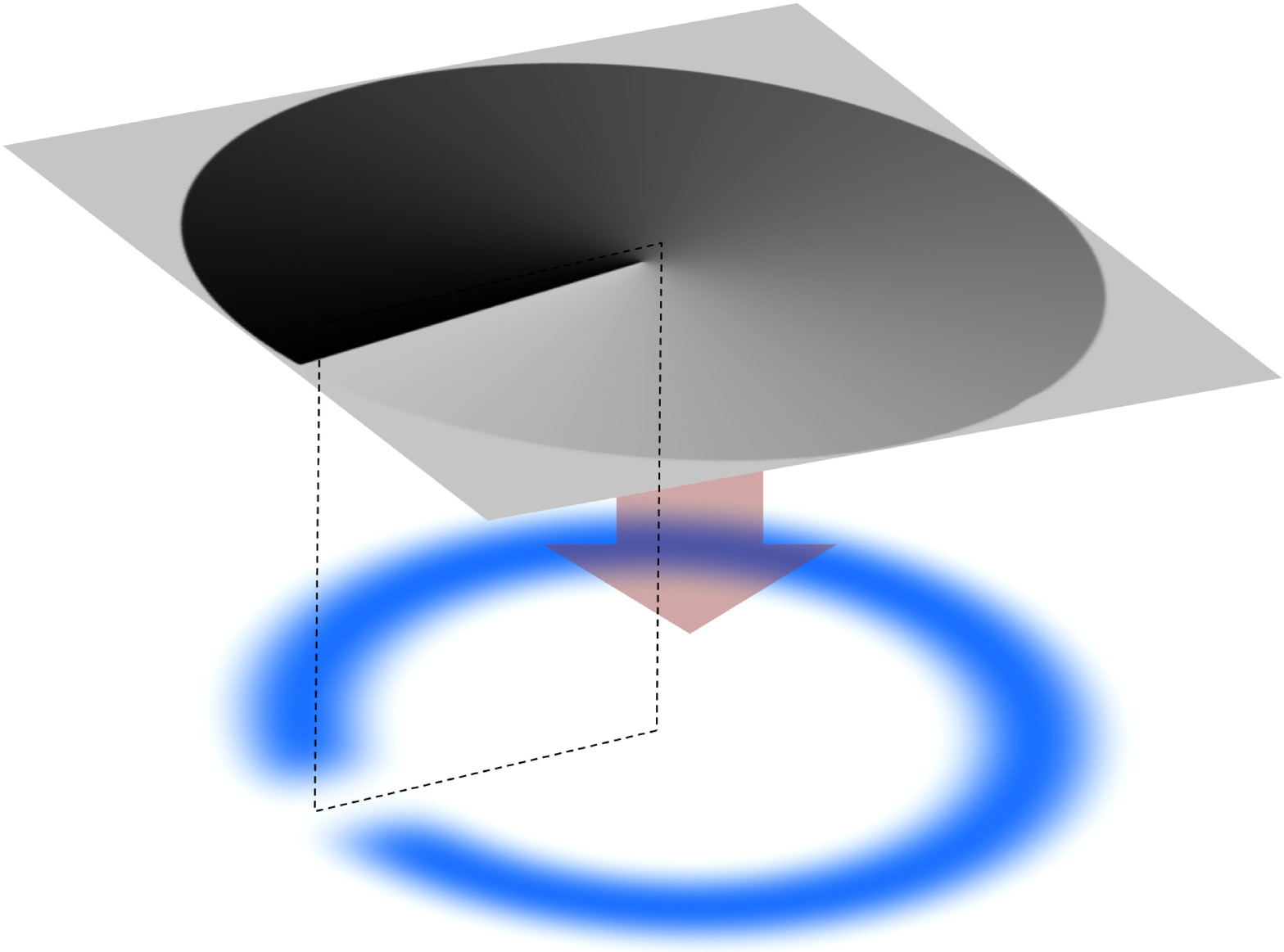}\\
		(b) & (c) & (d)
\end{tabular}
	\caption{SLM setup and generated intensity pattern: (a) Optical setup for an SLM used in the mask mode. The CCD camera is used here to observe the generated pattern. (b) Measured intensity pattern obtained from the SLM in mask mode for phase imprinting, with a beam waist of $200~\mu$m, imaged onto the CCD camera. (c) Azimuthal dependence of the normalized intensity profile across the red dotted circle, radius $50~\mu$m, of the obtained intensity profile in panel (b). (d) Proposed scheme to break the rotational symmetry by using a potential barrier produced with a laser beam and a subsequent phase imprinting. The edge of the intensity profile is aligned with the position of the barrier where the atomic density vanishes.}
	\label{Scheme}
\end{figure}

This method has been used to produce a soliton in an elongated condensate \cite{Burger1999,Denschlag2000}, with a stepwise intensity profile. In our experiment we want to set an annular condensate into rotation in a ring trap. As the superfluid velocity is related to the phase gradient of the condensate wave function, such motion can be described by a uniform phase gradient along the ring, of the form $\varphi(\theta)=\ell\theta$, where $\ell$ is the winding number and $\theta$ the azimuthal coordinate \cite{Dobrek1999}. To imprint such a phase we hence need to prepare an intensity pattern which is increasing linearly with the angle $\theta$ [see \fref{Scheme}(b)].

\subsection{\label{sec:holography}Tailoring the intensity pattern using an SLM}
As stated in the previous section, the desired light intensity pattern for the phase imprinting is a ``helix'' of intensity, with a linear dependence on the azimuthal angle $\theta$. Such a profile can be generated with a spatial light modulator, a device consisting of a matrix of pixels producing on an incident laser a computer-controlled local phase shift onto one of the polarization axes, the extraordinary axis \cite{Gaunt2012,Bowman2017}. The SLM can work in two modes, known as ``diffraction mode'' and ``mask mode''. In the diffraction mode, a light beam with a polarization parallel to the extraordinary axis of the SLM is sent on the device. The resulting pattern is located in the Fourier plane  and results from the diffraction on the phase grid programed on the SLM. The phase shift pattern to be programed has to be deduced from the desired pattern by running an inversion algorithm. By contrast, in the mask mode the polarization is aligned at 45$^\circ$ from the SLM neutral axes. The optical setup is thus prepared in a crossed polarizer analyzer configuration, with two polarizing beam splitters (PBSs). The desired pattern is programed directly on the SLM matrix, which changes the polarization accordingly. The target profile is obtained after filtering through the final PBS acting as an analyzer. While part of the incident power is lost through the other output of the PBS, this technique allows a more direct preparation of arbitrary light patterns and has been preferred in this work, as it achieves smoother profiles and can be fine-tuned using a passive feed back  \footnote{See supplemental material.}.

In order to imprint a given circulation, we have to shape the laser beam profile such that its intensity increases linearly with the azimuthal angle [see \fref{Scheme}(b)]. Starting from a Gaussian laser beam with a $1/e^2$ radius of $200~\mu$m, we generate such an intensity with an optical setup using an SLM (Hamamatsu X10468-07) in mask mode [see \fref{Scheme}(a)]. The useful pattern is imaged onto the atoms after filtering the useful polarization with a PBS. The peak intensity and the pulse duration can be adjusted such that the phase accumulated during the pulse peaks at an integer multiple of $2\pi$, thus preparing a rotating quantum gas with a well-defined winding number.
The pulse duration is set to 20~$\mu$s, much faster than the expected dynamics. A typical single-shot picture recorded on the CCD camera is shown in \fref{Scheme}(b).

\subsection{\label{sec:density}Overcoming the finite optical resolution}
In the ideal case the imprinted phase $\varphi$ should gradually increase with the azimuthal angle $\theta$ from $\varphi=0$ at $\theta=\theta_0$ to $\varphi=2\pi\ell$ at $\theta=\theta_0+2\pi$, for a targeted winding number $\ell\in\mathbb{Z}$, and be discontinuous at the starting angle $\theta_0$. This implies in turn an intensity profile with a discontinuity at $\theta=\theta_0$. Such a discontinuity in intensity, however, is not possible to produce in practice. The range in angle $\Delta\theta$ over which the intensity goes back to zero is set by the diffraction limit. In our optical setup this limitation is illustrated in \fref{Scheme}(c) in which the intensity at a fixed radius of 50~$\mu$m is plotted against the azimuthal angle, showing a range $\Delta\theta$ of $\sim$ 0.5~rad. This value depends on the radius at which the azimuthal profile is plotted, and the final resolution on the annular gas will thus depend on the ring radius. As a result of the rapid intensity decay on a finite range, the imprinted phase will induce a high atomic velocity in the direction opposite to the desired rotation, and the total angular momentum will vanish $\left\langle L_{z} \right\rangle =0$, in agreement with the results of Ref. \cite{Zheng2003}. Moreover, this phase imprint with a large local gradient induces high energy excitations in the sample, with velocities possibly larger that the critical velocity of the superfluid.

In order to overcome the issue related to the resolution limit, we propose to remove the atomic density in the region $\Delta\theta$, where the phase gradient is large but finite. This can be done by focusing a far-off-resonant blue-detuned light beam which repels the atoms and breaks the rotation invariance [see \fref{Scheme}(d)].

After phase imprinting,  in order to allow the rotation, the barrier needs to be removed, fast enough to prevent the quantum gas from getting reflected at the barrier. However we expect that an abrupt barrier removal will create excitations in the gas. The barrier removal time can thus be optimized. Another degree of freedom that we can adjust is the imprinted phase profile itself, which can also deviate slightly from a linear profile to compensate for the effect of the subsequent barrier removal. In the following we present a simulation of this transition from a broken rotational invariance to its restoration by numerically solving the Gross-Pitaevskii equation (GPE) for a condensate in a 2D ring trap. The goal of this calculation is, by analyzing the final state, to find the optimal barrier removal time and the optimal phase profile to be imprinted to reach the desired state.

\section{\label{sec:Simulations}GPE simulations}
We describe the dynamics of the trapped condensate with the mean-field model given by the Gross-Pitaevskii equation. We restrict the description to two dimensions, in the horizontal plane containing the ring trap. This applies directly to 2D annular quantum gases, which can be prepared in hybrid optical and adiabatic potentials \cite{Morizot2006}. We expect that our results will also be valid for three-dimensional quantum gases, as the dynamics will essentially occur within the ring plane.

\subsection{Initial state preparation}
\label{sec:initial_preparation}
The ring trap is described by a rotationally invariant, radial harmonic confinement with angular frequency $\omega_r$ and harmonic oscillator length $a_r=\sqrt{\hbar/(M\omega_r)}$, where $M$ is the atomic mass. From now on we will use dimensionless variables, scaled with the radial harmonic units, such that the unit of length is $a_r$, the unit of time is $\omega_r^{-1}$, and the unit of energy is $\hbar\omega_r$. The ring radius in these units is denoted $r_0$, and the ring trap potential simply reads: $V_{\rm ring}(r) = (r-r_0)^2/2$. With these units, the dimensionless 2D Gross-Pitaevskii equation in the ring, with a time-dependent barrier, reads
\begin{equation}
\begin{split} 
i  \frac{\partial \psi}{\partial t} = \left[ -\frac{1}{2} \nabla^{2} + V(r,\theta,t) + \tilde{g}N\left| \psi\right|^{2} -\mu  \right]\psi .
\end{split}
\label{eq:GPE}
\end{equation}
Here $\psi$ is the condensate wave function normalized to unity, $N$ is the atom number, $\tilde g$ is the 2D interaction strength \cite{Petrov2000a}, and $\mu$ is the chemical potential in units of $\hbar\omega_r$. The 2D trapping potential formed by the ring trap and the time-dependent barrier is given in polar coordinates $(r,\theta)$ by
 \begin{equation}\label{potential}
 \begin{split} 
V(r,\theta,t) = \frac{1}{2}\left(r-r_0\right) ^{2} +V_{B}(t) \,  e^{\displaystyle-\frac{(\theta-\theta_{B})^2}{2\sigma_{\theta}^2}},
 \end{split}
 \end{equation}
where $V_{B}(t)$ is the time-dependent height of the potential barrier in units of $\hbar\omega_r$, $\theta_{B}$ is the center of the barrier in the azimuthal coordinate, and $\sigma_{\theta}$ is the angular width of the barrier.

\begin{figure}[t]
\centering
	\includegraphics[height=0.28\columnwidth,clip=true]{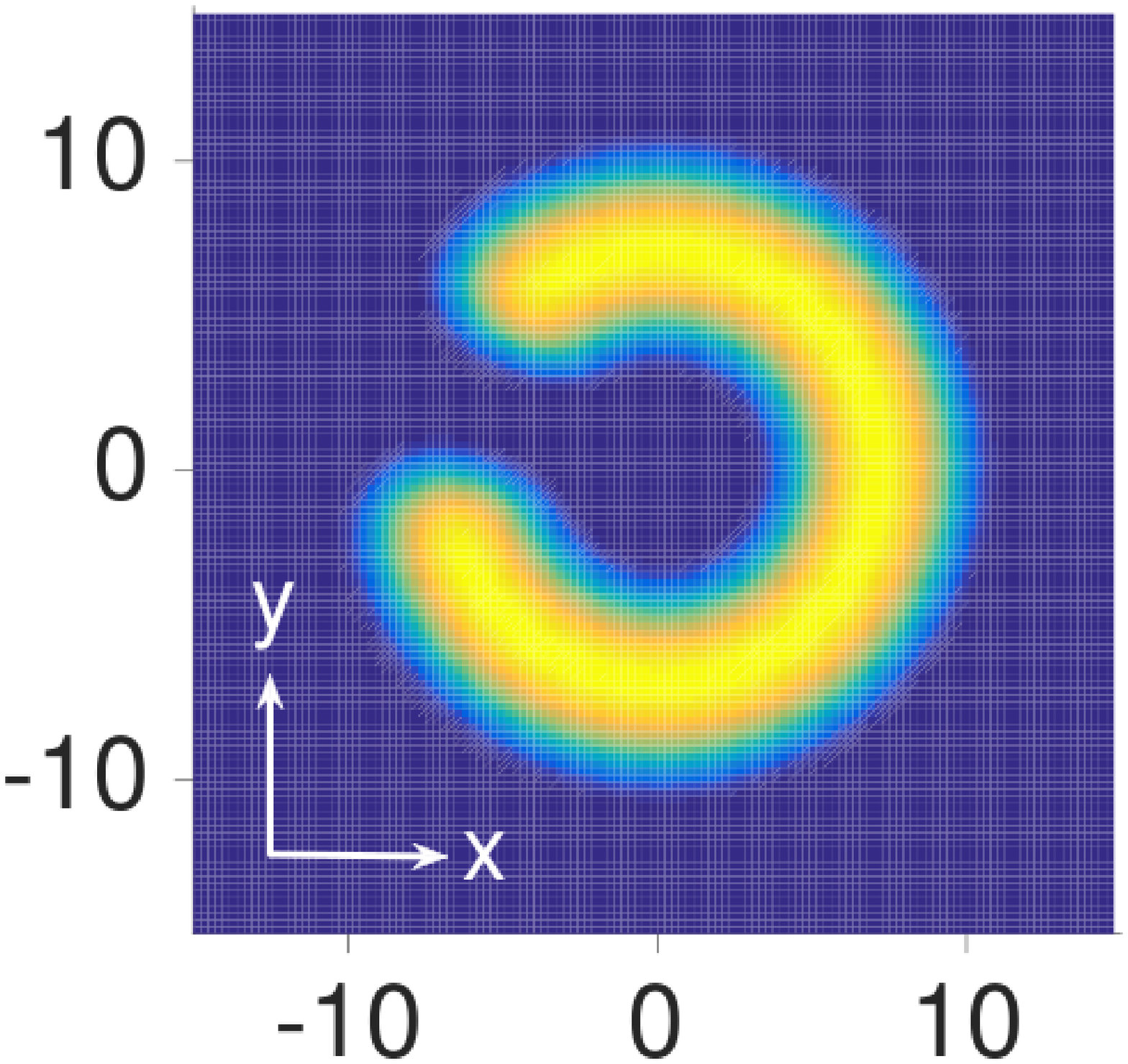}\includegraphics[height=0.28\columnwidth,clip=true]{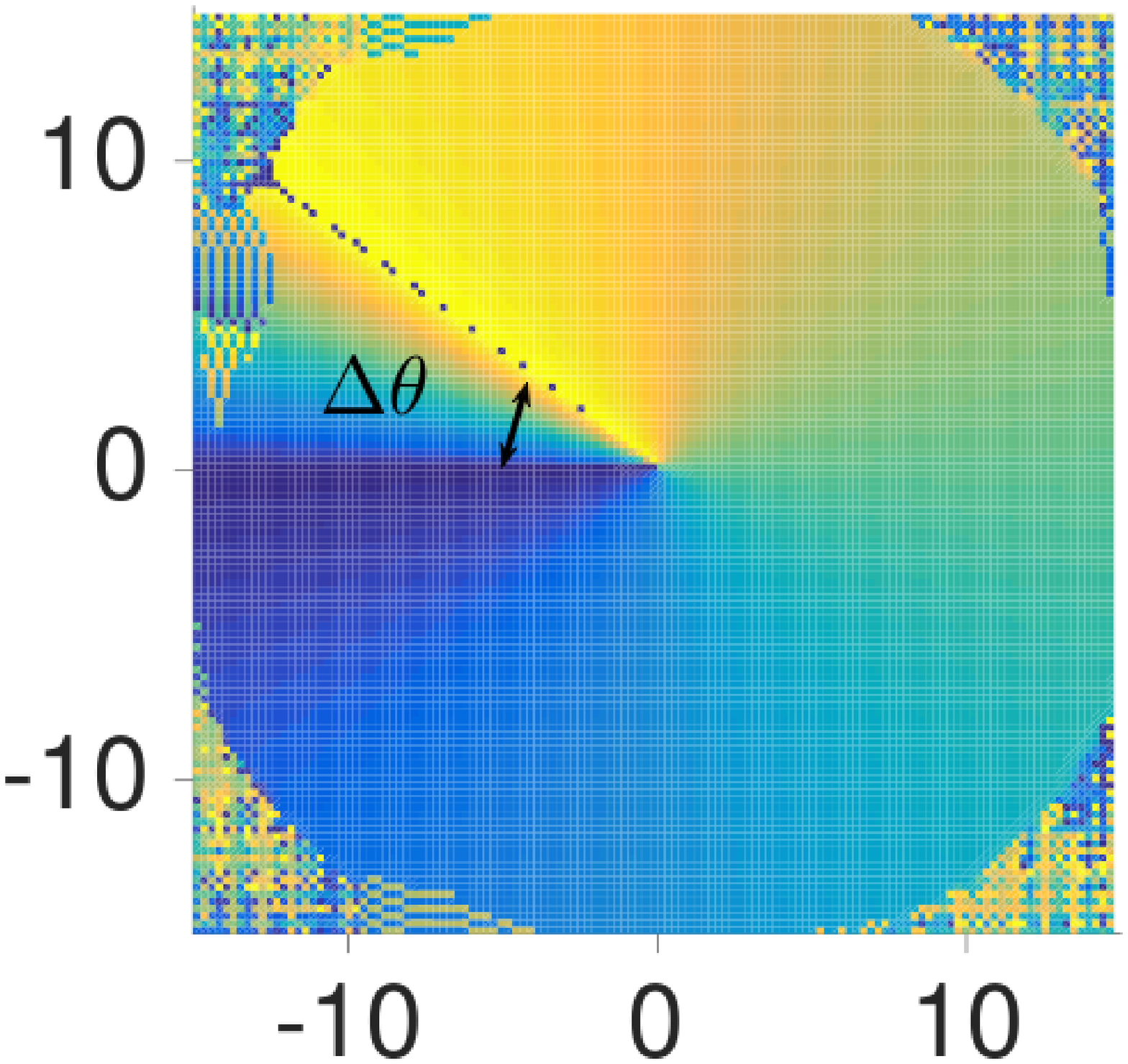}\includegraphics[height=0.28\columnwidth,clip=true]{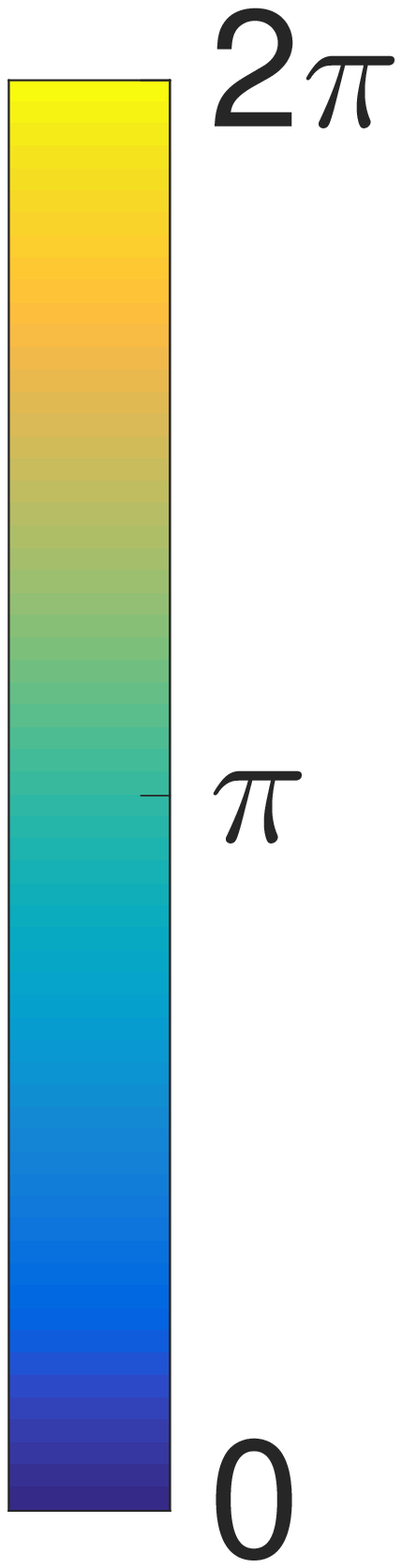}\includegraphics[height=0.28\columnwidth,clip=true]{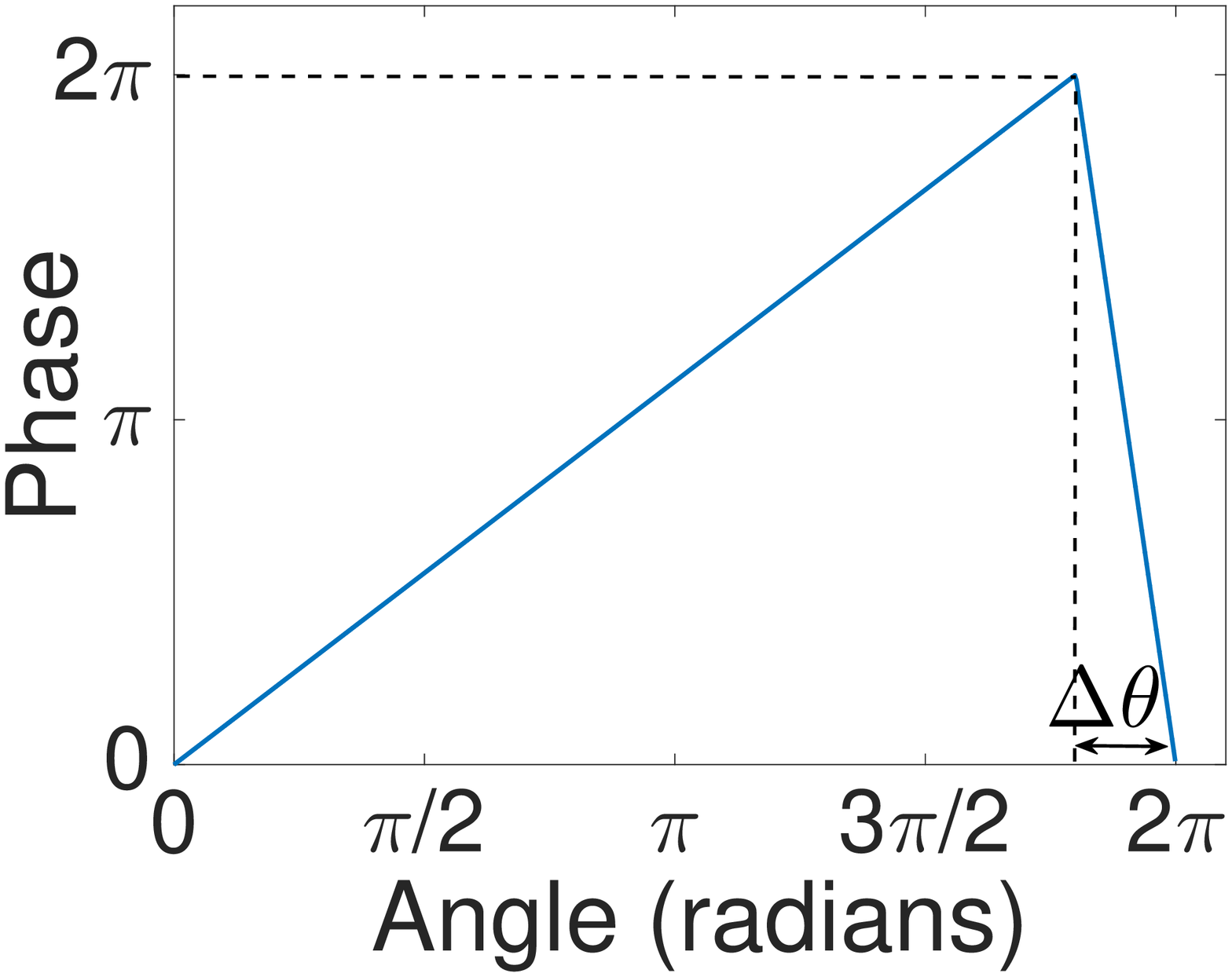}\\
	(a)\hspace{0.25\columnwidth} (b)\hspace{0.3\columnwidth} (c)
	\caption{ 
		(a) Ground-state density profile for broken rotational symmetry computed in the simulation; (b) phase profile imprinted on it to create an $\ell=1$ circulation state represented in two dimensions or (c) as a function of $\theta$. This is the starting point for the simulations in the presence of an initial barrier. The color scale for the phase profile is the same for all 2D phase plots.}
	\label{fig:InitialDensity}
\end{figure}

We use the split-step fast Fourier transform method~\cite{Bao2003} on a Cartesian square grid of 128 points in each direction. The grid size in dimensionless units is 30 and the trap ring radius is $r_0=7$. The coupling constant and the atom number are such that $\tilde{g}N=1000$. The initial ground state $\psi_0(r,\theta)$ is found by computing the evolution in imaginary time in the presence of the barrier, whose width is set to $\sigma_\theta=0.22$ and whose initial height is $V_B(t=0^-)=V_0=10$. The chemical potential with these figures is found to be $\mu=5.8$. The barrier width and height are chosen to allow a density drop larger than 80\% in the whole zone of width $\Delta\theta$ where the phase varies rapidly, such that the number of atoms affected by the sharp phase gradient remains very small. The presence of the barrier breaks the rotational symmetry of the ring as shown in \fref{fig:InitialDensity}(a).

\begin{figure}[t]
		\includegraphics[width=\columnwidth,clip=true]{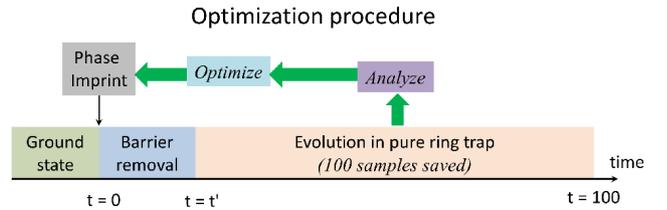}
		\caption{Optimization algorithm to transfer the system into the desired state. The time evolution is divided into three main sequences: ground-state computation, barrier removal, and evolution in the smooth ring potential. In the third sequence the wave function is statistically analyzed and optimization is done on the barrier removal ramp to reach the desired output state.\label{fig:optimization}}
\end{figure}

On the initial condensate prepared in the ground state of GPE in the presence of the barrier $\psi_0$, an helicoidal phase profile is imprinted instantaneously. In order to take into account the practical limitations induced by the finite optical resolution limit, we model the imprinted phase $\varphi(\theta)$ with a piecewise linear function \cite{Zheng2003}, increasing from $0$ to $2\pi\ell$ over the range $2\pi-\Delta\theta$ and then going back to zero over the small angle $\Delta\theta=2\pi/10$, slightly above the experimentally measured value of $0.5$~rad, as shown in \fref{fig:InitialDensity}(c). The position of the barrier is chosen to match this rapid phase change, such that $\theta_B=2\pi-\Delta\theta/2$. The phase imprint process is very fast as compared to the atomic motion, and in the simulation the initial wave function $\psi_0(r,\theta)$ is simply multiplied by the imprinted phase factor: $\psi(r,\theta,t=0^+)=\psi_{0}(r,\theta)\exp\left[  i\varphi(\theta)\right
 ]$. Thanks to the annular shape of the gas, this phase profile does not lead to any discontinuity of the wave function in the center. This wave function is then evolved in real time through Eq.~\eqref{eq:GPE}, which describes the barrier removal and the subsequent evolution in the ring-shaped potential alone (see \fref{fig:optimization}).

\subsection{\label{sec:simulationSystem} Optimum barrier removal and phase profile}
After the initial state preparation, the numerical simulation is divided into two more steps (see \fref{fig:optimization}). In a first step, the barrier is removed, with a linear ramp in intensity, right after phase imprinting. The evolution is thus computed in the presence of a time-varying potential. In a second step, after the barrier has been removed completely, which occurs at a time $t=t'$, the wave function is evolved until time $t_{\rm end}=100$ (in dimensionless units) in the static ring potential $V_{\rm ring}$ alone. For our later analysis, we save 100 frames of the evolution after the barrier removal, between $t'$ and $t_{\rm end}$, and extract the angular momentum for each (see below). We then optimize the barrier ramp and the imprinted phase pattern from these results.

In order to analyze the result in terms of angular momentum transfer, we expand the total angular momentum into states of pure circulation such that $L_{z}(t)=\sum_{m} K_m(t)m$ in units of $\hbar$, where $m\in\mathbb{Z}$ specifies the circulation state. $K_m$ is thus the total population in the states with a given angular momentum $m\hbar$. We compute these expansions for all of the 100 frames we extracted, by taking a Fourier transform in the azimuthal space of the radially averaged wave function. Preparing a persistent current with the winding number $\ell$ would correspond to the case where $K_\ell=1$ and $K_m=0$ for all $m\neq \ell$. We use the following cost function $\mathcal{C}$ to optimize the barrier removal ramp and the phase pattern:
\begin{equation}\label{eq:cost}
\begin{split} 
\mathcal{C} = \left\langle  \left( 1-K_\ell \right) ^{4}\right\rangle_{20<t<t_{\rm end}}.
\end{split}
\end{equation}

\begin{figure}[t]
\centering
(a)\includegraphics[height=0.4\columnwidth,clip=true]{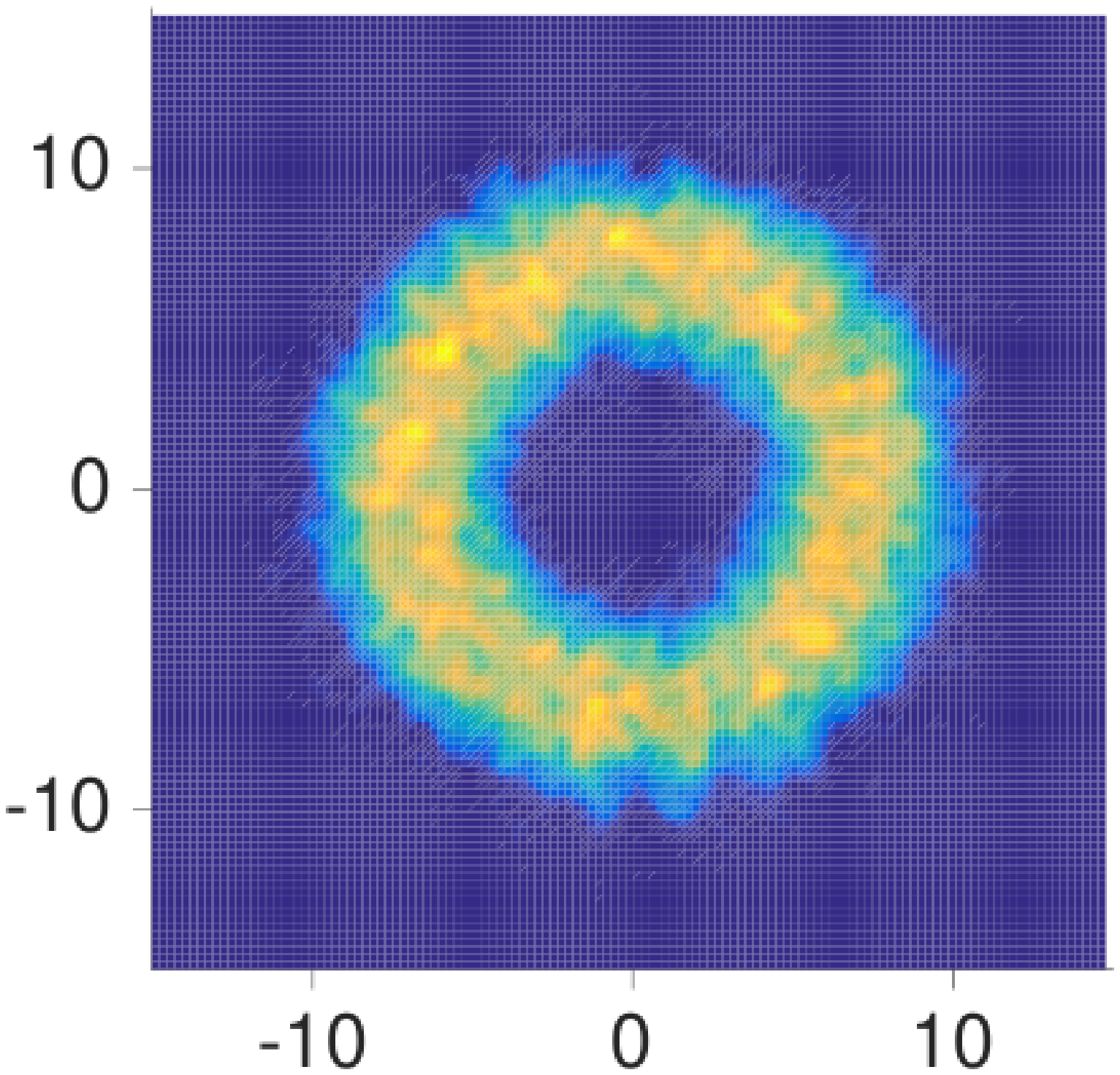}\hspace{0.05\columnwidth}(b)\includegraphics[height=0.4\columnwidth,clip=true]{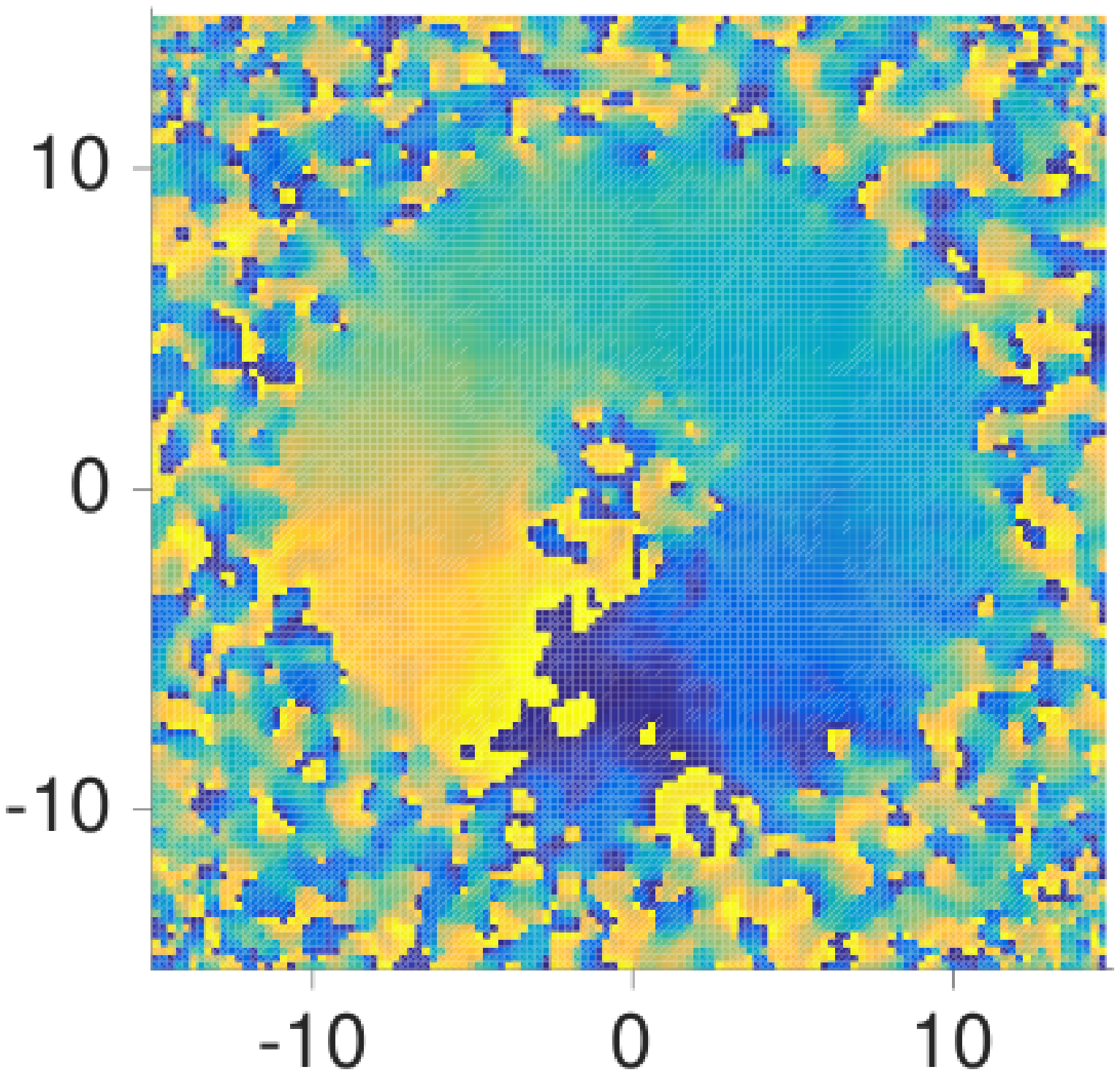}\\
(c)\includegraphics[width=\columnwidth,clip=true]{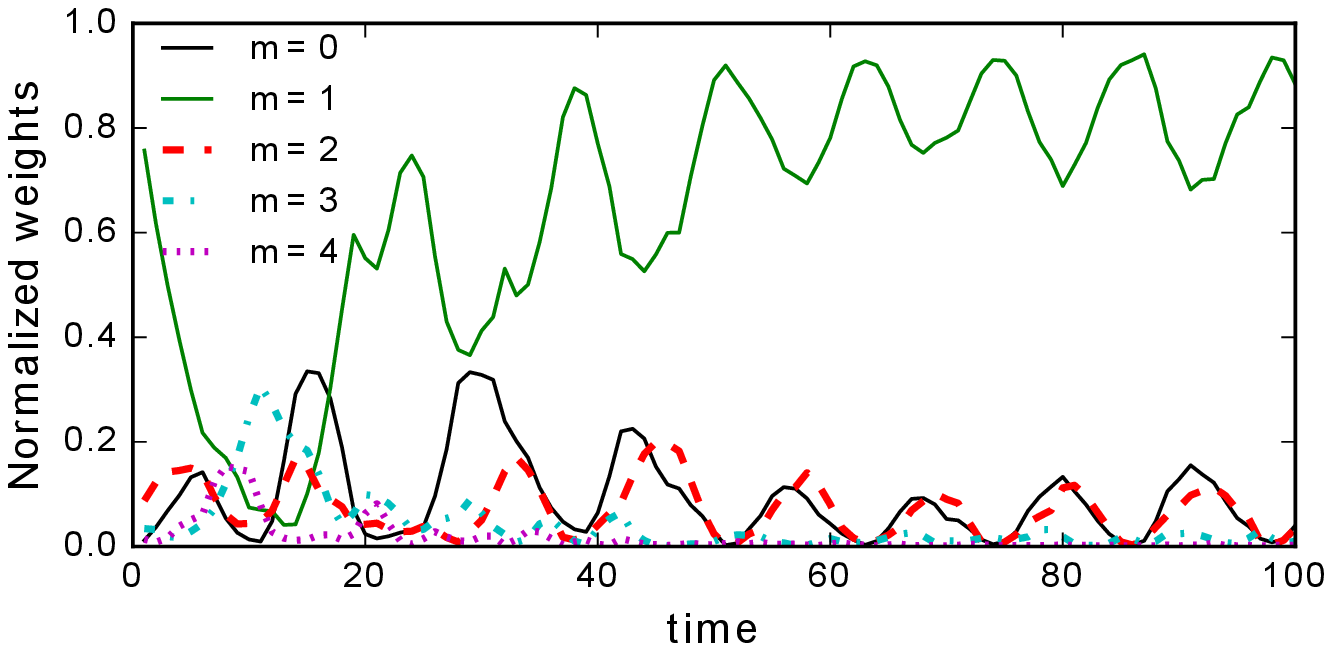}
	\caption{(a) Density and (b) phase profiles after the phase imprint of \fref{fig:InitialDensity}(c) and for an optimized barrier removal time of $t'=0.5$, after the full evolution in the ring trap ($t=100$). Phase color scale is as in \fref{fig:InitialDensity}. (c) Time evolution of the population in different $m$ circulation states.\label{fig:opt_barrier}}
\end{figure}

Using this figure of merit, we first optimize the barrier removal ramp with the fixed phase profile shown in \fref{fig:InitialDensity}(c) for $\ell=1$. We find that the minimum value of $\mathcal{C}$ is obtained for a rather short optimal removal time of $t'=0.5$ (see \fref{fig:opt_barrier}). The final circulation $\ell=1$ is prepared, and the final state is free from vortex excitations in the bulk [see Figs.~\ref{fig:opt_barrier}(a) and \ref{fig:opt_barrier}(b)]. However, large oscillations in the populations in the different $m$ circulation states are still present [see \fref{fig:opt_barrier}(c)]. While this optimization allows us to get rid of vortices in the bulk, we find that the cost function after the optimization is not very much reduced as compared to an abrupt removal ($t'=0$) \cite{Note1}. In fact the critical parameter to optimize is instead the phase profile imprinted, as we show below. In order to reduce the number of parameters to be optimized, we present in the following the results of the phase profile optimization obtained with an abrupt removal of the barrier ($t'=0$). The results obtained with a nonzero value of $t'$ have been checked to be similar, as soon as phase profile optimization is performed.

We then optimize the phase imprint profile itself, for an abrupt removal ($t'=0$). The idea of this approach is to include in the phase imprint an additional term to compensate for the acceleration that the barrier removal will induce. Using the SLM we can generate any desired intensity distribution with a resolution only limited by the diffraction limit of the optical system.  The phase profile is thus written in the interval $(0, 2\pi -\Delta\theta)$ in the form of a truncated Fourier series whose coefficients are to be optimized for a given target winding number $\ell$:
\begin{equation}\label{Imprint_eqn}
\varphi(\theta) = \ell\theta + \sum_{n=1}^{n_{\rm max}} \left[C_{n}\cos \left(\frac{n\theta}{2}\right)+  S_{n}\sin \left(\frac{n\theta}{2}\right) \right].
\end{equation}
In the interval $(2\pi -\Delta\theta,2\pi)$, it decreases linearly back to its value $\varphi(0)$ at $\theta=0$ .

We optimize the contribution of the Fourier components $ C_{n}$ and $ S_{n}$ using the steepest gradient descent \footnote{The whole code, including the GPE part, is run using Matlab. The optimization routine uses the \emph{fmincon} function. The differential step in the function is set to be 0.1 to avoid the convergence towards local minima. The full optimization is performed on a desktop computer and takes about 10 hours for a given $\ell$.}. We set the frequency cutoff to $n_{\rm max}=4$ for $\ell=1$ and $2$ or to $n_{\rm max}=6$ for $\ell=3$ as going higher does not give significant improvement in the optimization. In any case, the finite optical resolution would limit $n_{\rm max}$ to $n_{\rm max}=10$ to be consistent with our choice of $\Delta\theta=2\pi/10$.

\begin{figure}[t]
	\centering
	{\includegraphics[width=\columnwidth,clip=true] {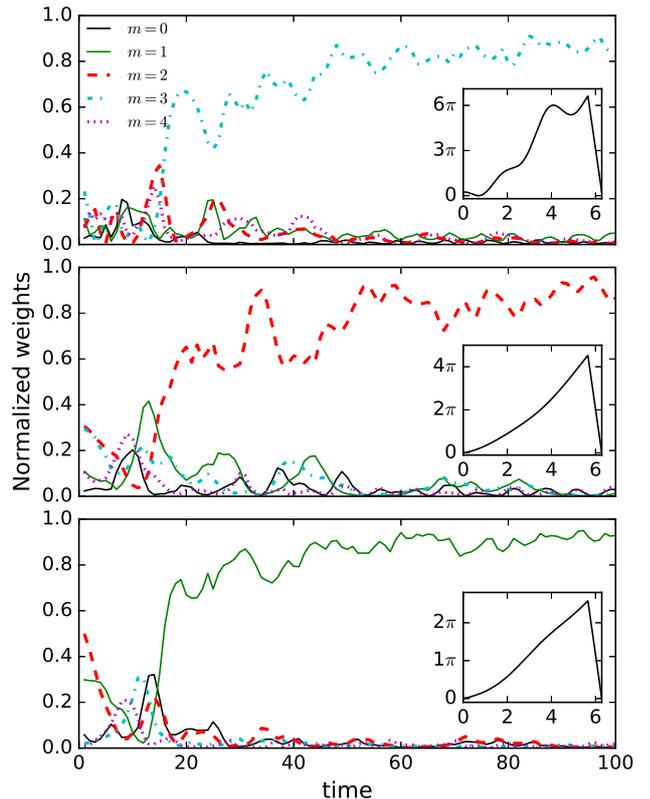}}
	\caption{ 
		The evolution of the population in different $m$ circulation states using a nonlinear imprint obtained from optimization as described in the main text for $\ell=3$ (top panel), $\ell=2$ (middle panel) and $\ell=1$ (bottom panel) states. The insets show the imprints that lead to the time evolution in the main figures.}
	\label{fig:optimizationResults}
\end{figure}
\subsection{\label{sec:results} Results}
\Fref{fig:optimizationResults} shows the results of the optimization of the phase profile for three target states: $\ell=1$, $\ell=2$, and $\ell=3$. The insets show the nonlinear imprints $\varphi(\theta)$ obtained through the optimization algorithm. The total phase difference imprinted is close to $2\pi\times\ell$, although a bit larger. These imprints can be easily produced using an SLM. After imprinting these phase patterns and at the end of the time evolution, the population $K_\ell$ in the target state is $K_\ell\sim 0.9$, with some fluctuations. The excess energy added in this process relative to the energy of the ideal circulating states is 0.2768, 0.3137, and 0.5441 (in units of $\hbar\omega_r$) for the cases of $\ell=1$, $\ell=2$, and $\ell=3$, respectively. They are at least ten times smaller than the chemical potential and will scarcely increase the temperature of the system in typical experimental conditions. This excess energy can be removed through evaporation.

\begin{figure}[t]
\centering
\begin{tabular}{ccc}
	\raisebox{0.15\columnwidth}{\rotatebox{90}{$t=0$}} & \includegraphics[height=0.4\columnwidth,clip=true]{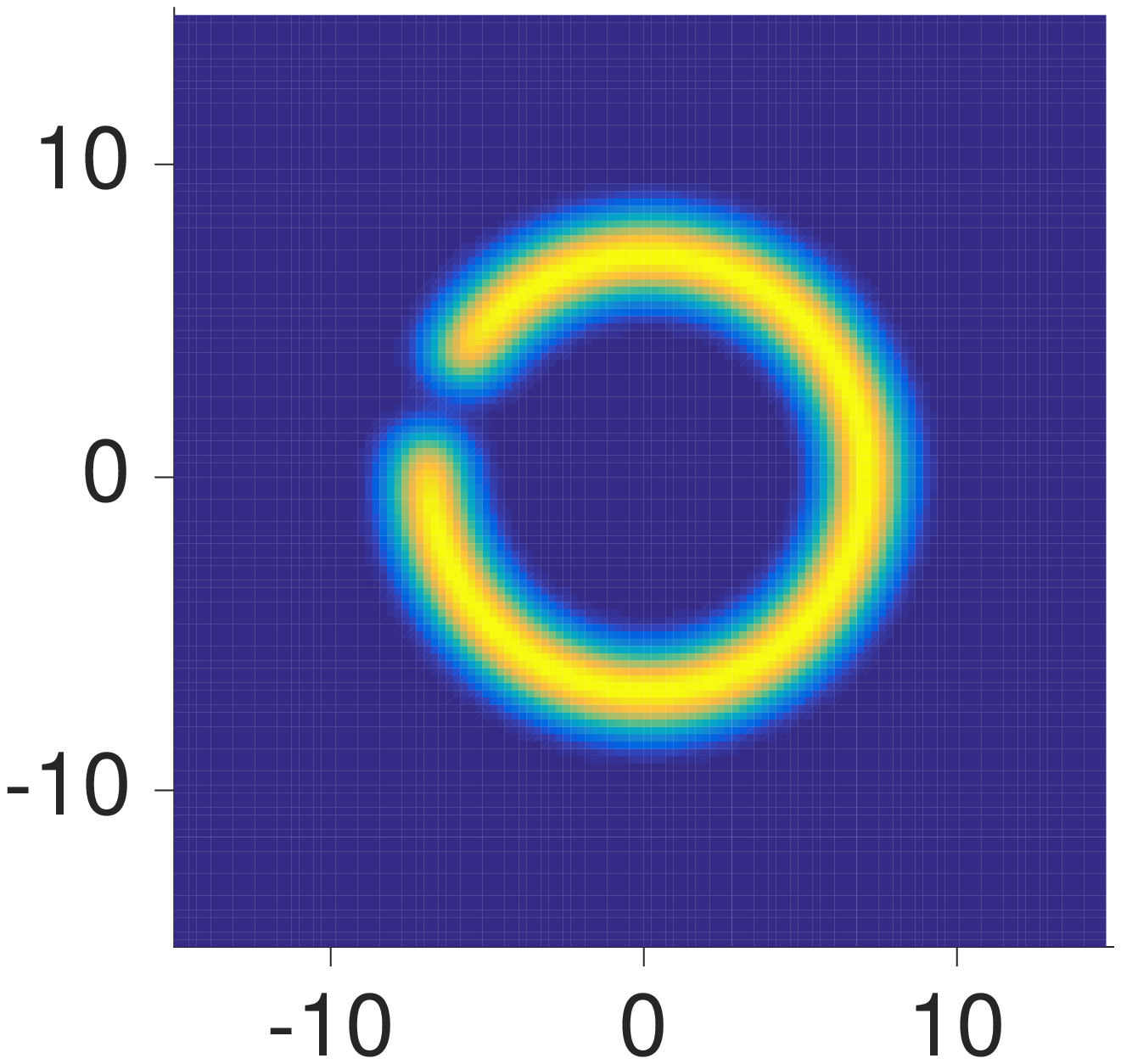} & \includegraphics[height=0.4\columnwidth,clip=true]{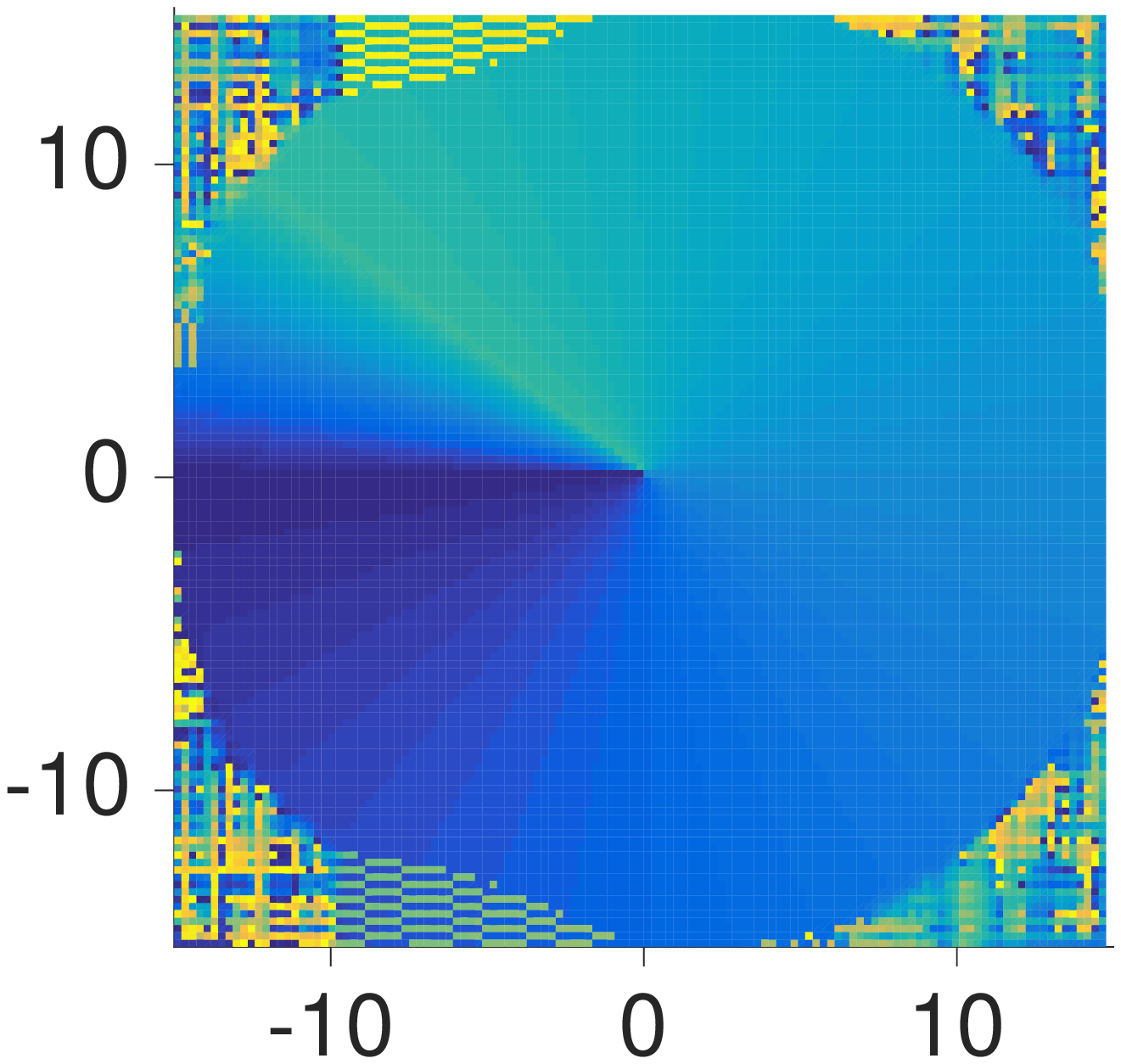}\\
	\raisebox{0.15\columnwidth}{\rotatebox{90}{$t=100$}} & \includegraphics[height=0.4\columnwidth,clip=true]{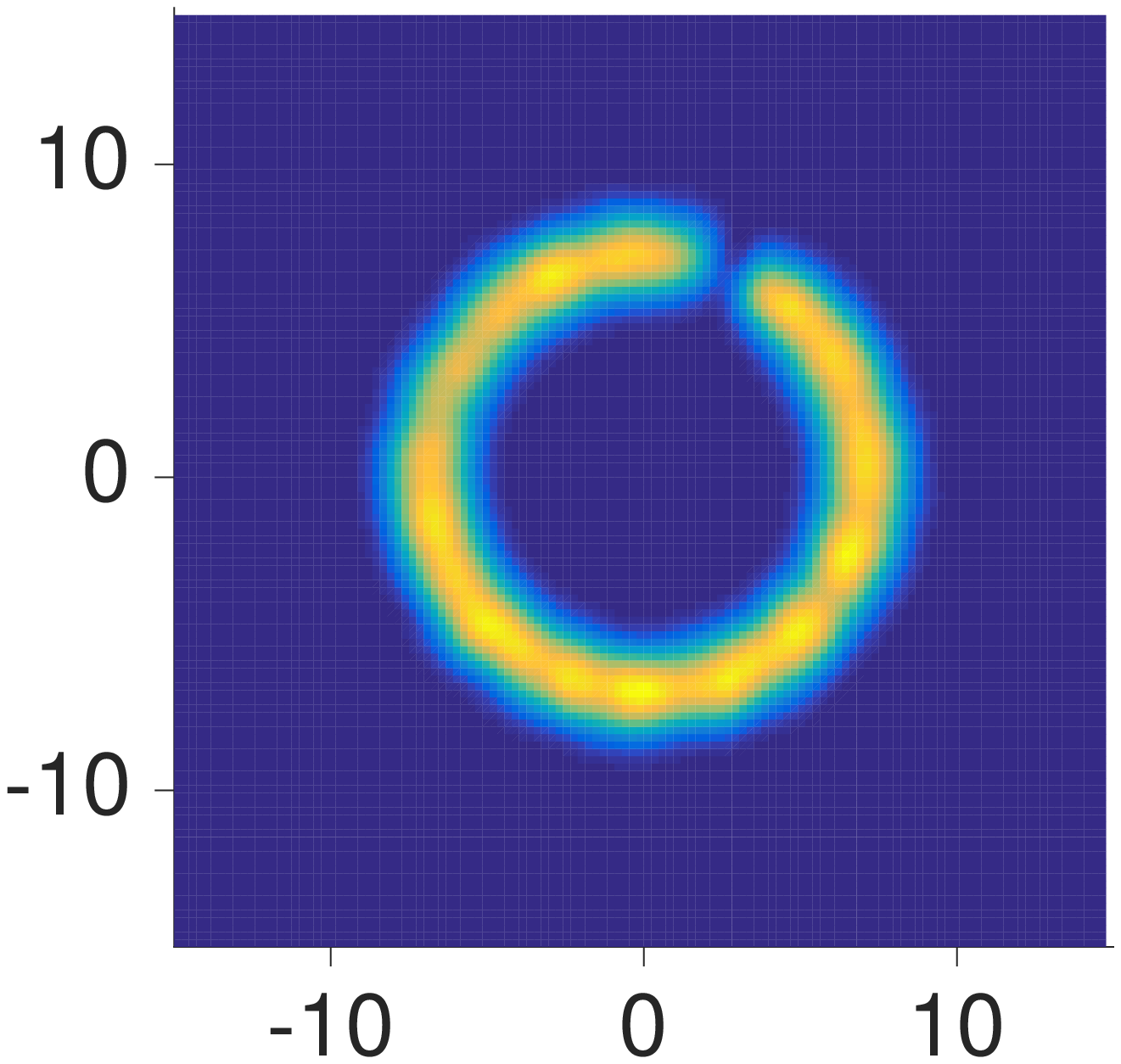} & \raisebox{-0.005\columnwidth}{\includegraphics[height=0.4\columnwidth,clip=true]{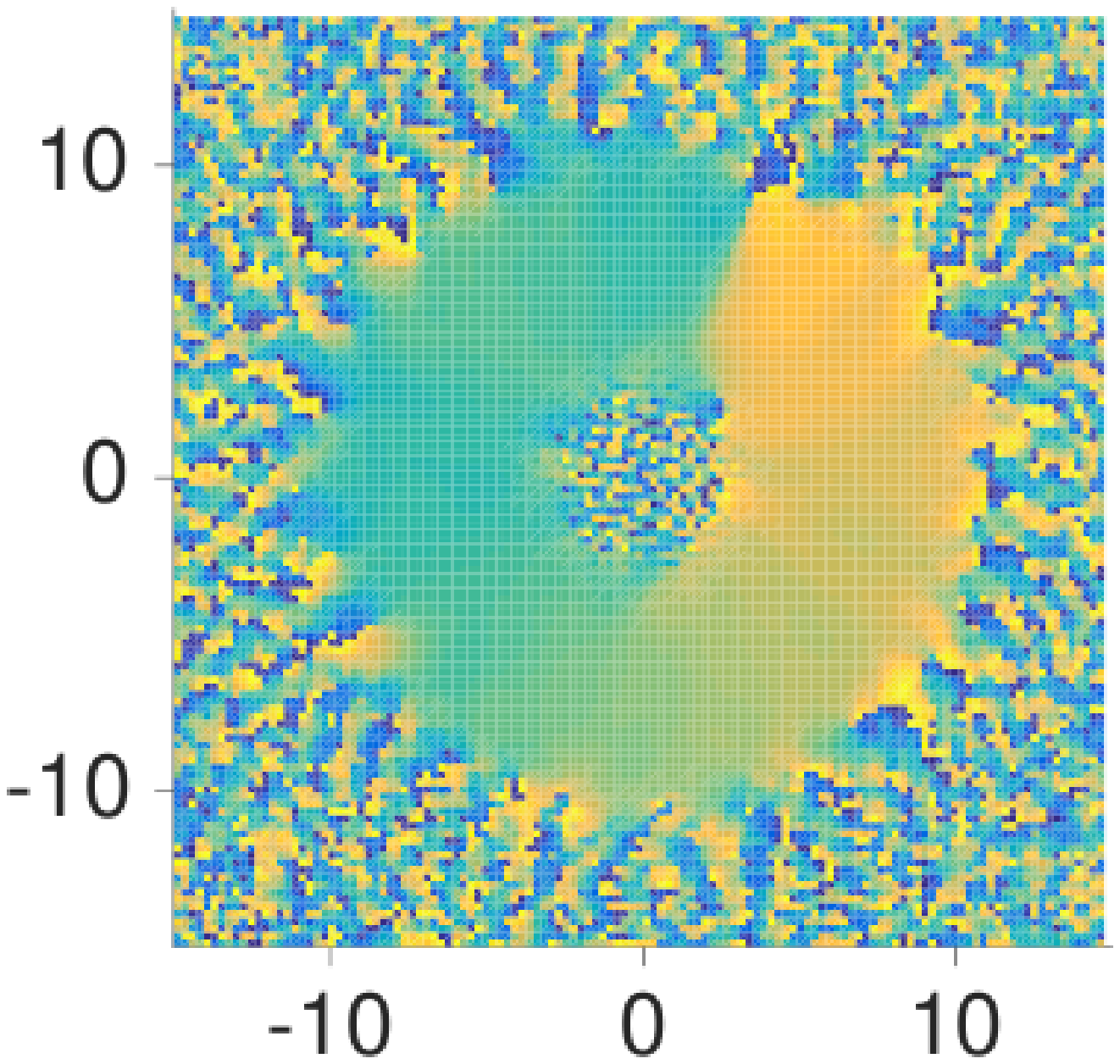}}\hspace{0.11\columnwidth}~\\
	&$|\psi|^2$ & $\varphi$
	\end{tabular}
		\caption{ 
		Generation of a moving gray soliton using phase imprinting and barrier removal on a 1D annular gas. A phase jump of $\pi$ is imprinted at $t=0$ (top row) and it stabilizes to a phase jump of $0.63\pi$ with some density perturbations moving in the opposite direction (bottom row). Phase color scale is as in \fref{fig:InitialDensity}.\label{fig:Soliton}}
\end{figure}

We also simulated this method of phase imprinting for the preparation of a gray soliton in a quasi-1D ring. This approach has been used before for an elongated condensate confined in a cigar-shaped trap \cite{Burger1999,Denschlag2000}. The dynamics of a gray soliton depends on the amplitude of the phase jump \cite{Frantzeskakis2010} and corresponds to a static, dark soliton, when the phase jump reaches $\pi$. In principle the phase jump of the soliton can be tuned to any value in a straightforward way using the SLM. By adjusting this phase we can create a dark soliton, stationary with respect to the background fluid, or a moving soliton at any subsonic velocity. As solitons are stable only in 1D systems, we perform these simulations using $\tilde{g}N=100$, which reduces the chemical potential below the radial confinement energy, and we compute the ground state of the quasi-1D annular gas in the presence of a barrier of width $\sigma_\theta=0.07$ and height $V_0=2.5$. We then imprint the phase with an imperfect phase jump aligned with the barrier position and remove the barrier abruptly. \Fref{fig:Soliton} shows a gray soliton created by imprinting a phase jump of $\Delta\varphi=\pi$ that stabilizes to a phase change of $0.63\pi$ across the density dip, which rotates around the ring trap. We end up with a gray soliton because the width of the barrier (limited by the imperfect phase profile) is larger than the intrinsic soliton width, set by the healing length. We observe that adding a barrier significantly helps to get rid of the density waves reported in Ref.~\cite{Burger1999} using the same method of phase imprinting. We note that by optimizing the shape of the imprinted phase it should be possible to control the final phase jump and hence the soliton properties.

\begin{figure}[t]
	\includegraphics[width=\columnwidth,clip=true]{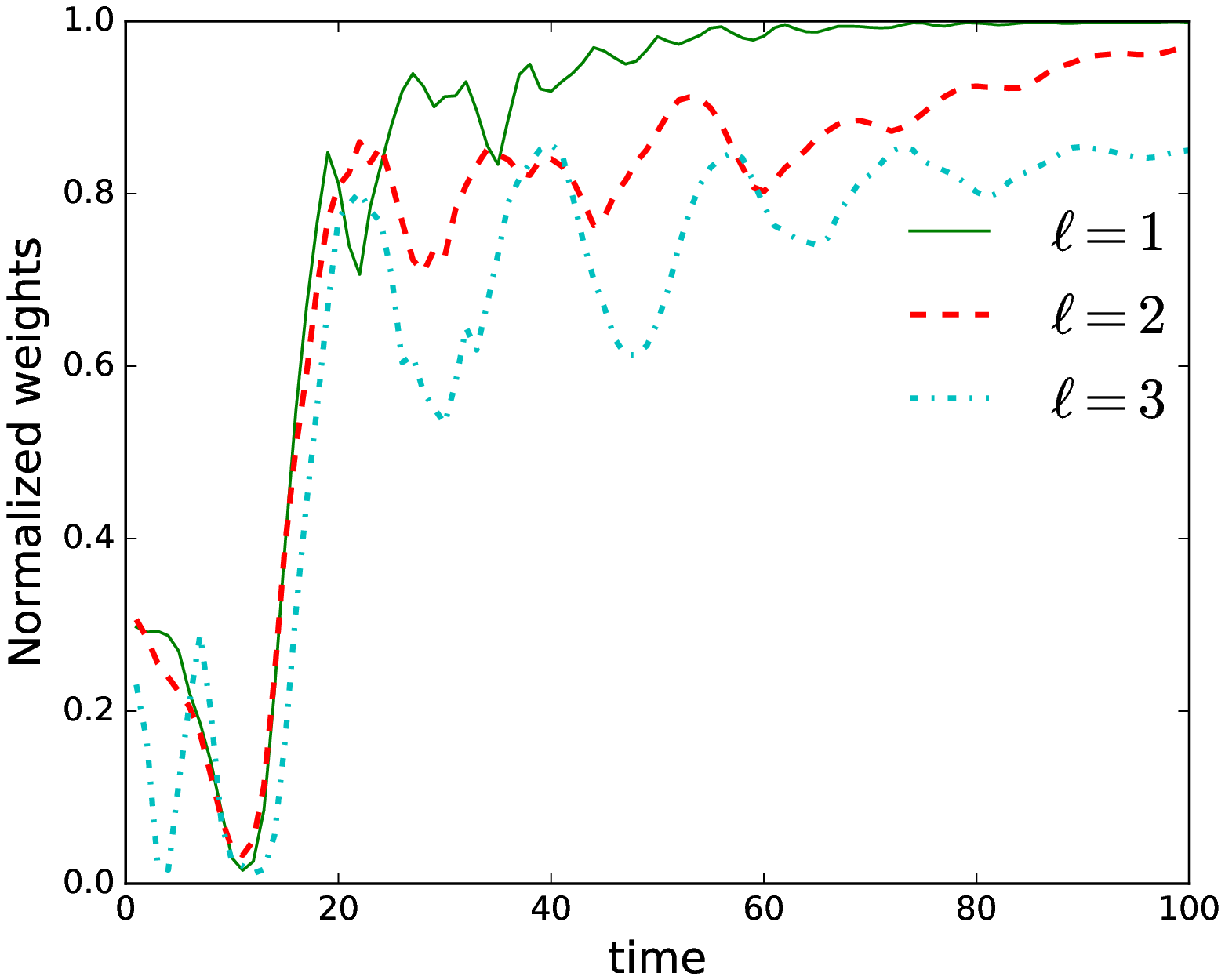}\\[-0.38\columnwidth]
	\hspace*{0.25\columnwidth}\includegraphics[height=0.25\columnwidth,clip=true]{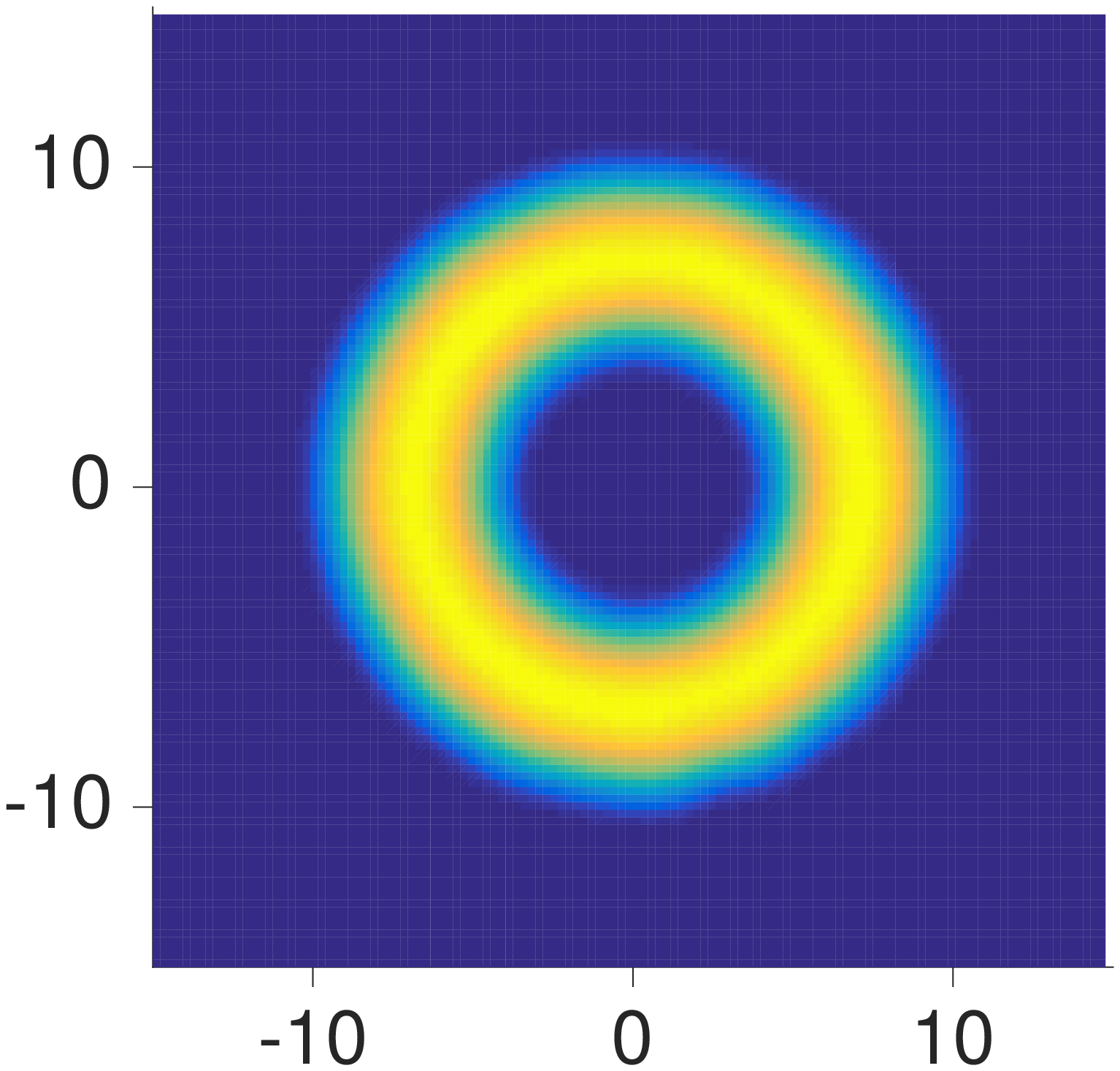}\hspace{-0.01\columnwidth}
	\includegraphics[height=0.25\columnwidth,clip=true]{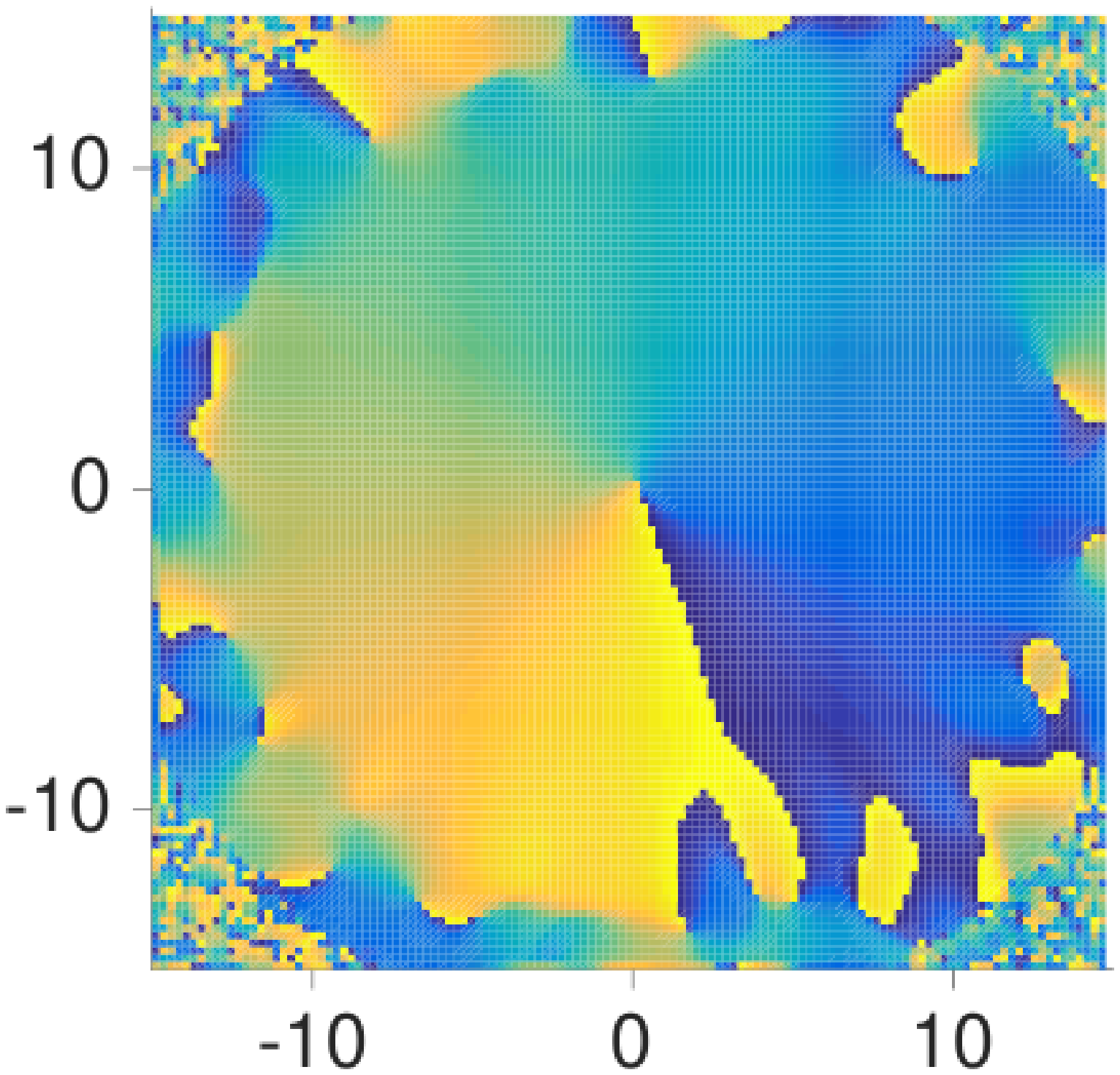}\\[0.05\columnwidth]
~
	\caption{Evolution of the population in the $m=\ell$ states after a phase imprint optimized for a given target value $\ell$. Damping is introduced by setting $\gamma=0.01$. The final density (left inset) and phase (right inset) profiles at the end of evolution are shown for the case $\ell=1$. Phase color scale is as in \fref{fig:InitialDensity}.\label{fig:LcompDamped}}
\end{figure}

\section{\label{sec:Dissipation} Including damping}
The Gross-Pitaevskii equation describes the evolution of the quantum gas dynamics in the absence of losses or damping. In order to take into account the effect of dissipation in the experiment, due to atom loss or to the finite trap depth leading to evaporation, we include a small imaginary part to the time evolution, using the dimensionless parameter $\gamma$, which describes a phenomenological damping \cite{Choi1998}.

The dimensionless GPE including the phenomenological damping term is now given by
\begin{equation}
\begin{split} 
i  \frac{\partial \psi}{\partial t} = \left( 1-i\gamma\right) \left[ -\frac{1}{2} \nabla^{2} + V(r,\theta,t) + \tilde{g}N\left| \psi\right|^{2} -\mu  \right]\psi.
\end{split}
\label{eq:GPEdiss}
\end{equation}

This phenomenological approach with $\gamma>0$ can be used to simulate the damping of the excitations when evaporation is on and to find the metastable state the system converges to. For the damping coefficient $\gamma$ = 0.01, also used in Ref. \cite{Eckel2014a}, the atomic state converges towards a stable circulation state. We have checked that the choice of $\gamma$ does not influence much the final state but rather the rate at which it is reached. The simulation including dissipation is run in the three cases $\ell=1$, $\ell=2$ and $\ell=3$, analogous to the one presented in \fref{fig:optimizationResults} in the absence of dissipation. The results for the evolution of the population in the $m$ states are shown in \fref{fig:LcompDamped} together with the final density and phase profile in the insets. The dissipation helps to remove the remaining fluctuations of population in the various $m$ states and helps the system to converge to a state where nearly all the population is concentrated in states with an angular momentum quantum number $m=\ell$. The role of the barrier is nonetheless essential and we checked that a mere phase imprint without the density depletion does not converge efficiently to a state with a non zero circulation, even if damping is introduced.

\section{\label{sec:Conclusions}Conclusions}
In summary, we have presented a practical method to set an annular quantum gas into a given circulation state using phase imprinting. In order to overcome the practical diffraction limit arising from the tailored light profile, we have simulated the behavior of the condensate after a realistic phase imprint in the presence of a barrier. Our simulations show that it is possible to prepare a given circulation state as well as other designed dynamical states like solitons by carefully engineering the phase imprint. While optimizing the barrier removal time alone allows us to suppress bulk vortex excitations, we find that the optimization of the phase pattern is crucial to achieve a high fidelity in the preparation of the target circulation. For example we reach a population in the $\ell=1$ circulation states of $K_\ell=0.9988$ for an optimized phase profile when damping is introduced (see \fref{fig:LcompDamped}).

The phase imprinting method is also fast as compared to the adiabatic rotating barrier method. Following the protocol of Ref.~\cite{Eckel2014a} we simulated the preparation of an $\ell=1$ circulation state, by rotating a barrier for a full round trip along the annulus at the frequency $\Omega=1/r_0^2$ in dimensionless units, which takes a time $t_{\rm stirr}=2\pi/\Omega=2\pi r_0^2\simeq 300$ for our parameters. With this slow stirring protocol (as compared to the fast phase imprinting) the $\ell=1$ state is achieved with a fidelity above 0.9. Interestingly the transfer of circulation during the stirring process involves oscillations between the different $m$ states, reminiscent of what happens for the linear phase profile [see \fref{fig:opt_barrier}(c)], with a slightly longer period. When the stirring time is reduced by a factor of 2 to $t_{\rm stirr}=150$, these oscillations are more pronounced. The population $K_1$ in the target state still oscillates around 0.85 long after the end of the stirring time, up to the end of the simulation at $t=250$, such that the effective preparation time is not improved. The main improvement of our phase imprint optimization has been to reduce these oscillations, achieving faster convergence to the desired state (see \fref{fig:optimizationResults}). It would be interesting to study whether an optimization of the rotating barrier protocol could allow one to prepare well-defined circulation states faster than the slow timescale $t_{\rm stirr}=2\pi/\Omega$.

In a future work it would be very interesting to extend the phase imprint optimization method presented here to the precise control of soliton creation, in particular to create multiple solitons with well defined relative velocities, which would give access to the study of solitonic collisions \cite{Nguyen2014,Jezek2016}.

\acknowledgments
We thank Laurence Pruvost and Bruno Viaris for assistance with the SLM in the early stage of the experiment, and Paolo Pedri for helpful discussions on the numerical simulations. We acknowledge financial support from the ANR project SuperRing (Grant No. ANR-15-CE30-0012) and from the R\'egion Ile-de-France in the framework of DIM ``des atomes froids aux nanosciences,'' project PESR, and of DIM SIRTEQ (Science et Ing\'enierie en R\'egion \^Ile-de-France pour les Technologies Quantiques), project DyABoG.

\clearpage

\begin{center}
{\large\textbf{Supplemental material}}
\end{center}
\setcounter{section}{0}
\section{Intensity pattern generation using the SLM\label{sec:SLM}}
An SLM can be used in diffraction mode or in mask mode. Several techniques exist to generate the desired pattern by diffraction at infinity \cite{Pasienski2008,Gaunt2012, Bowman2017}. Nevertheless, we prefer to use the mask method in which the desired pattern is the conjugated image of the SLM plane by an optical system, after polarization analysis (see section II.B of the main text). This allows a straightforward control of the intensity, pixel by pixel. Moreover, we have implemented a feedback loop by imaging the SLM's plane onto a camera so that each pixel of the SLM corresponds to a group of pixels of the camera, to correct the intensity iteratively, pixel by pixel. At the cost of rejecting optical power, the measured pattern converges towards any arbitrary target pattern. This technique yields patterns with good fidelity in less than 10 iterations, does not require complex calculation nor careful calibration of the SLM, and allows to correct for the spatial defects of the incident laser beam.

\begin{figure}[h]
\centering
\begin{tabular}{cc}
\includegraphics[height=.35\columnwidth]{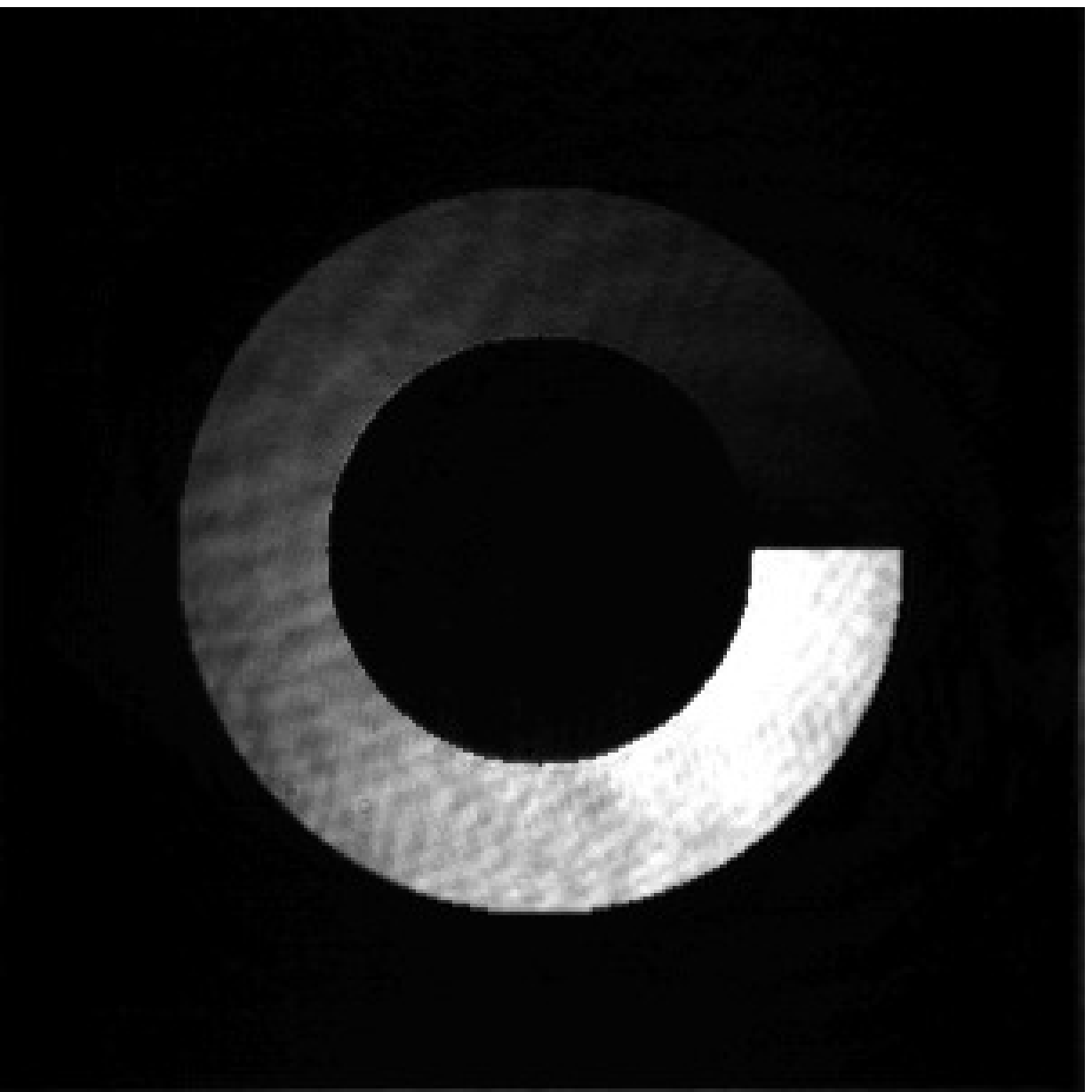} & \includegraphics[height=.35\columnwidth]{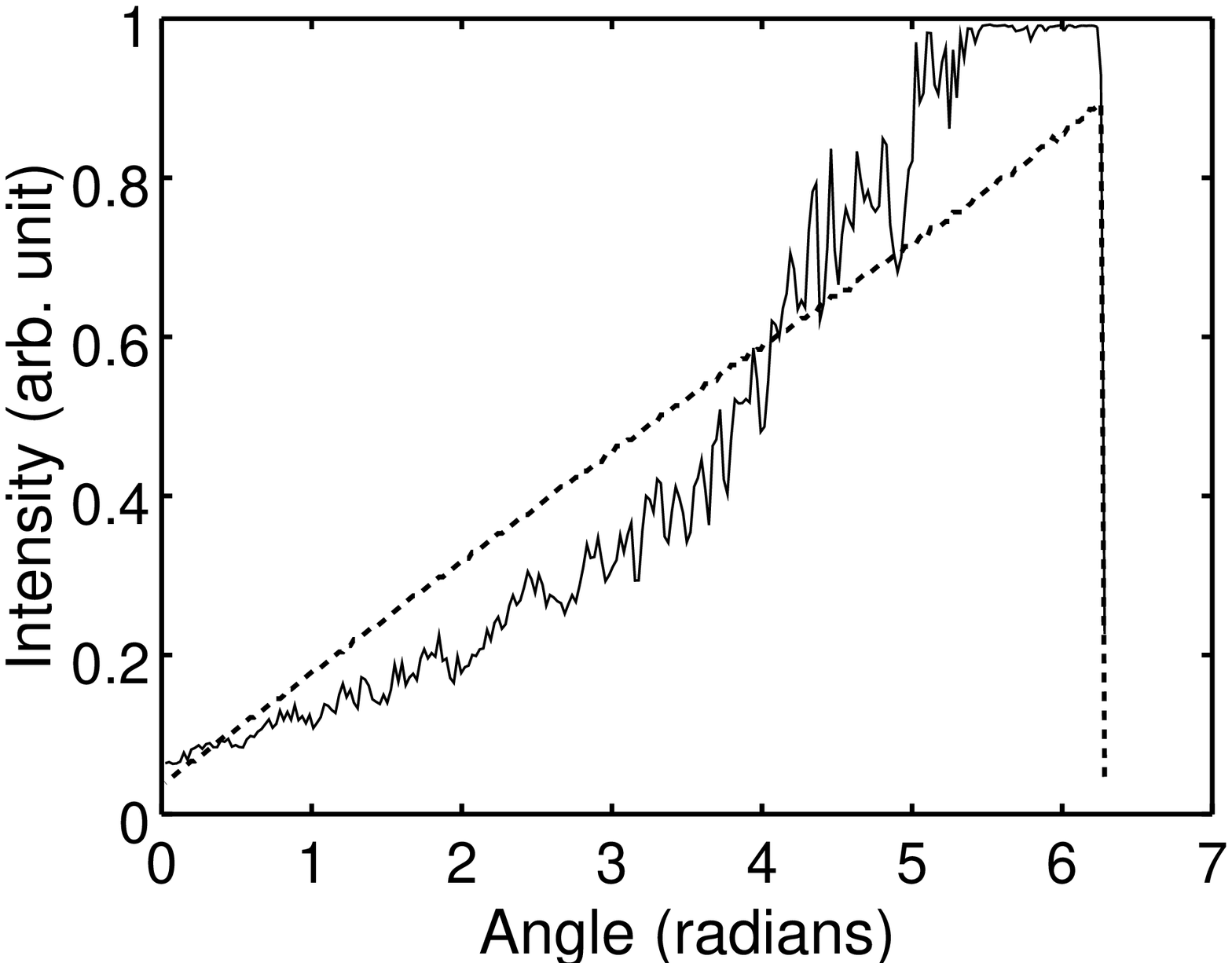}\\
	(a) & (b)\\
\includegraphics[height=.35\columnwidth]{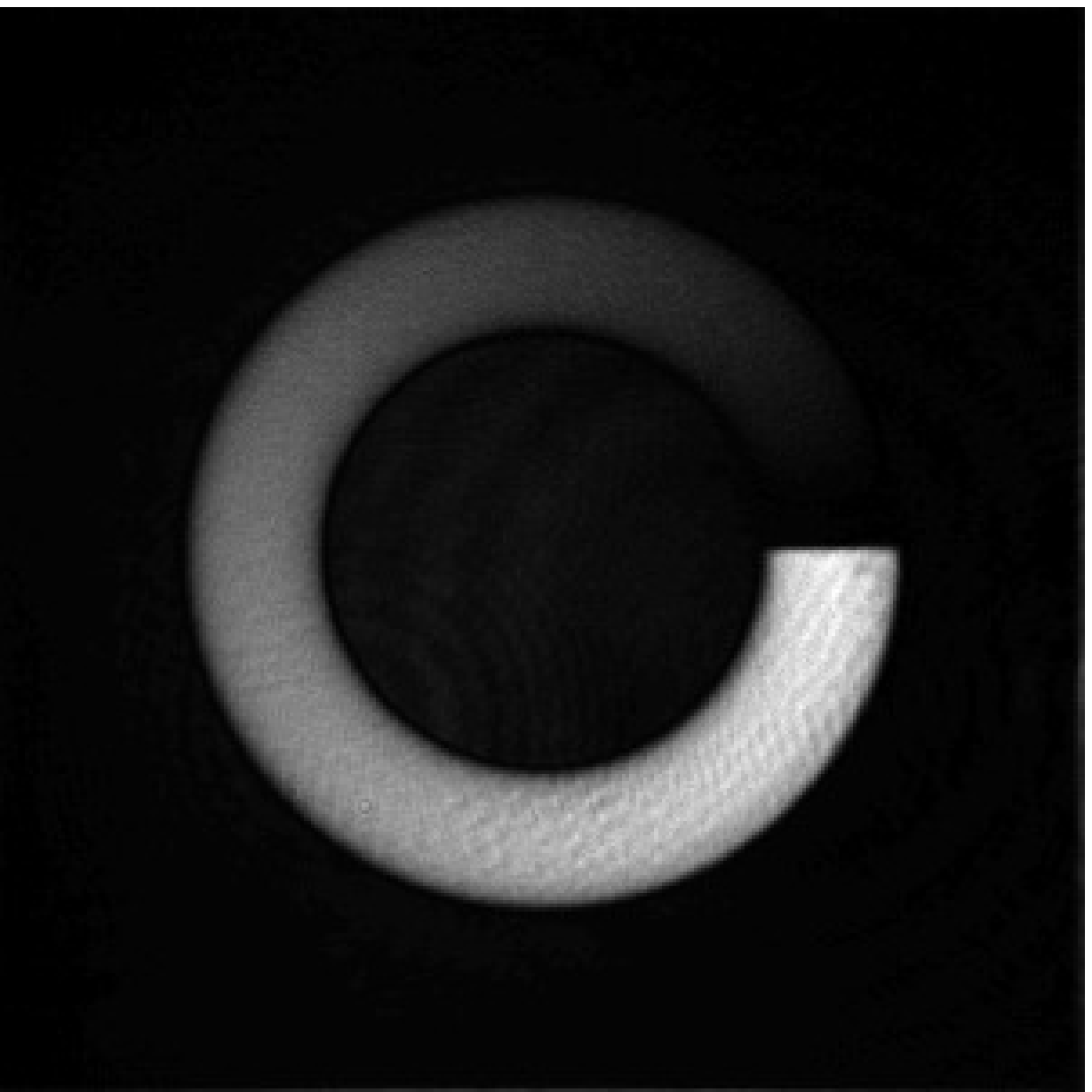} & \includegraphics[height=.35\columnwidth]{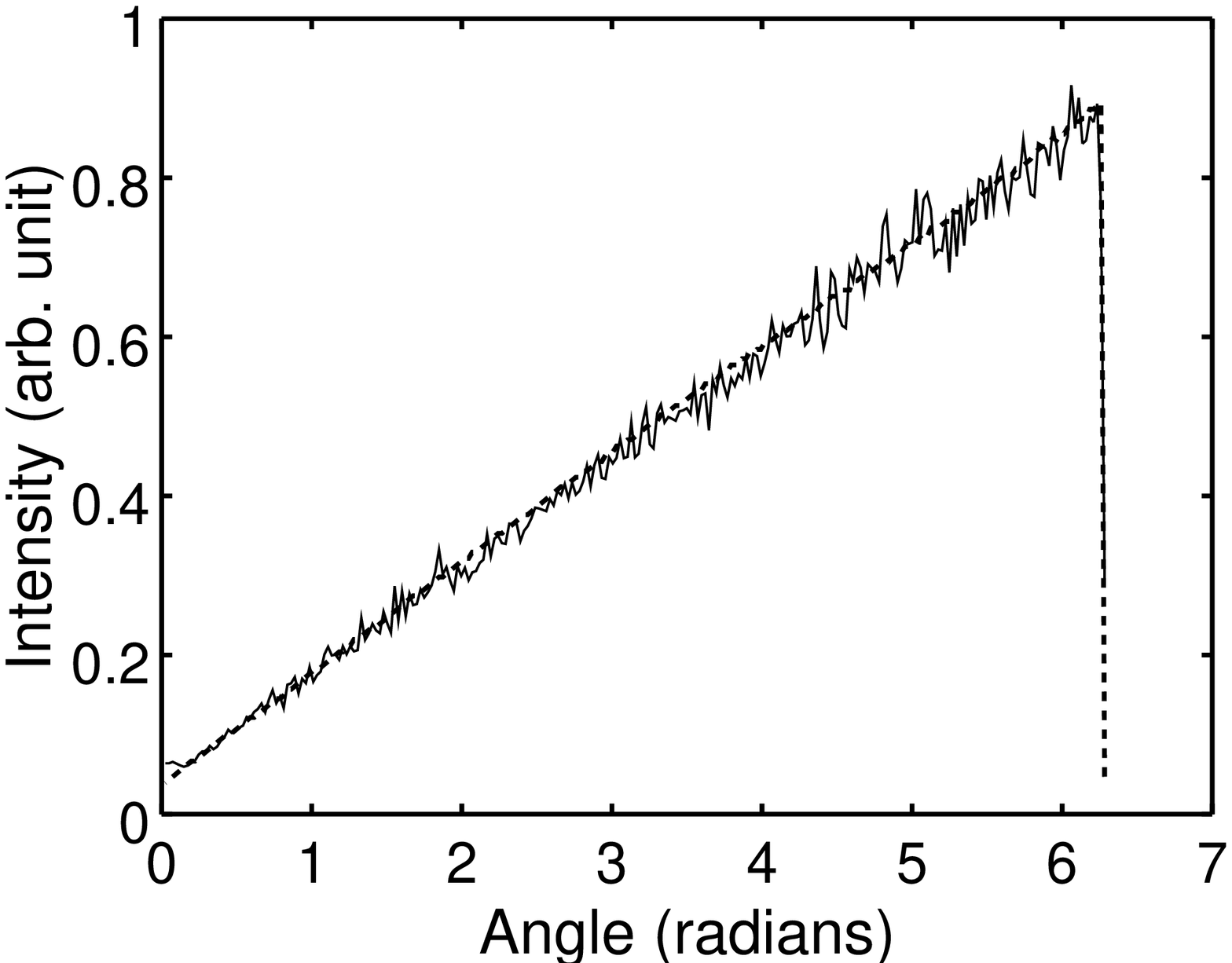}\\
	(c) & (d)
\end{tabular}
	\caption{Intensity patterns generated in mask mode (a) before and (c) after feedback. Also shown are the measured azimuthal profiles of the normalized intensity cut at a radius of 1.76~mm for the (b) initial and (d) final measured patterns. The dotted line is the target profile.}
	\label{fig:IntensityProfiles}
\end{figure}

Beyond the helix of intensity along the azimuthal angle combined with a radial Gaussian profile, see Fig. 1(b) of the main paper, it is possible to realize an helix of intensity with a flat radial profile and with the light limited to an area adapted to the annular gas. \Fref{fig:IntensityProfiles} shows such an intensity pattern, measured before and after feedback, as well as the normalized intensity along a given radius. We note that this pattern is produced with a large ring radius of $1.76$~mm, for which the diffraction limit leads to a very small intensity decay region with $\Delta\theta\sim 9$~mrad, hardly visible on \fref{fig:IntensityProfiles}(d). The rms deviation computed over the annular shape is 34\% for the initial profile and 6.45\% for the profile measured after 6 feedback steps. About 9\% of the total incident laser power on the SLM is transferred into the final pattern of \fref{fig:IntensityProfiles}(c). These values compare well with the experimental figures of diffraction techniques found in the literature.

\begin{figure}[b!]
	\centering
	{\includegraphics[width=1.0\columnwidth] {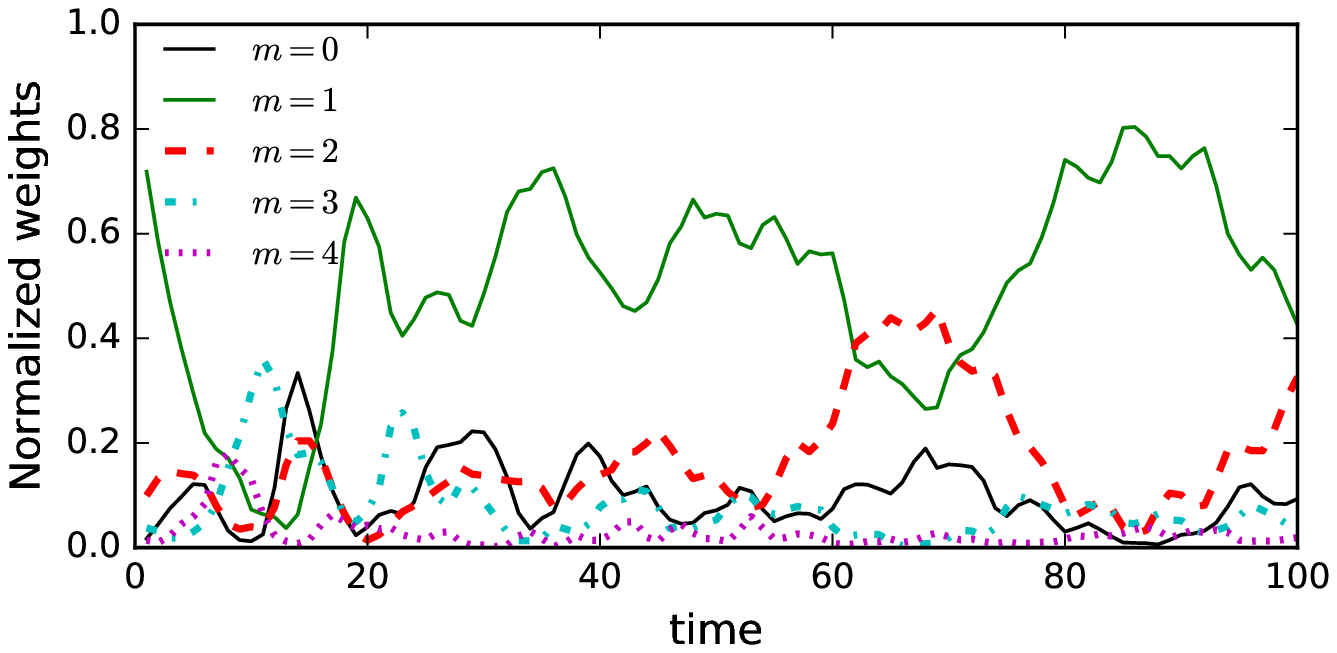}}
	\caption{Evolution of the population in the different $m$ circulation states for a linear phase imprint and an abrupt removal of barrier.}
	\label{LAbrupt}
\end{figure}

\section{Barrier Removal time}
As discussed in the main text, for a fixed phase imprint function, barrier removal time has an impact on the convergence to the target state. Figure \ref{LAbrupt} shows the evolution of population in different $m$ states when a linear phase for $\ell=1$ is imprinted and the barrier is removed abruptly. This unoptimized case gives a cost (as defined in main text) of $\mathcal{C}=0.06$ compared to the best case shown in the main text when the ramp time is $t'=0.5$ giving a cost of $\mathcal{C}=0.016$. However when the phase imprint is optimized for both of these cases we reduce to similar values in the final cost. In general, a poorly chosen barrier removal time can be compensated by an optimized phase profile except for very large barrier removal times when dipolar oscillations starts to play a role and the total angular momentum decreases considerably.

\end{document}